\documentclass[10pt,a4paper,twocolumn,aps,prb,floatfix,superscriptaddress,showpacs,showkeys]{revtex4-1}
\usepackage[english]{babel}
\usepackage[intlimits]{amsmath}
\usepackage{amssymb,graphicx}

\let\vec=\mathbf
\renewcommand\Re{\operatorname{Re}}

\begin{document}

\title{Theory of strong-field injection and control of photocurrent in dielectrics and\\
wide bandgap semiconductors}

\author{S.~Yu.~Kruchinin}
\email{stanislav.kruchinin@mpq.mpg.de}
\affiliation{Max-Planck-Institut f\"ur Quantenoptik, Hans-Kopfermann-Str.~1, D-85748 Garching, Germany}

\author{M.~Korbman}
\affiliation{Max-Planck-Institut f\"ur Quantenoptik, Hans-Kopfermann-Str.~1, D-85748 Garching, Germany}

\author{V.~S.~Yakovlev}
\email{vladislav.yakovlev@mpq.mpg.de}
\affiliation{Max-Planck-Institut f\"ur Quantenoptik, Hans-Kopfermann-Str.~1, D-85748 Garching, Germany}
\affiliation{Ludwig-Maximilians-Universit\"at, Am Coulombwall~1, 85748 Garching, Germany}

\begin{abstract}
We propose a theory of optically-induced currents in dielectrics and wide-gap semiconductors exposed to a non-resonant ultrashort laser pulse with a stabilized carrier-envelope phase.
In order to describe strong-field electron dynamics, equations for density matrix have been solved self-consistently with equations for the macroscopic electric field inside the medium, which we model by a one-dimensional potential.
We provide a detailed analysis of physically important quantities (band populations, macroscopic polarization, and transferred charge), which reveals that carrier-envelope phase control of the electric current can be interpreted as a result of quantum-mechanical interference of multiphoton excitation channels.
Our numerical results are in good agreement with experimental data.
\end{abstract}

\pacs{78.20.Bh, 72.80.Sk, 77.22.Ej, 42.50.Hz}

\keywords{%
dielectric, semiconductor, few-cycle pulse, strong-field physics, ultrafast optics, carrier-envelope phase, coherent control
}
\maketitle

\section{Introduction}\label{s:intro}

The time it takes electrons in a solid to respond to an external electric field is on the order of a fraction of a femtosecond.
Due to recent progress in ultrafast and attosecond science, it is now possible to perform attosecond-scale time-resolved measurements of extremely nonlinear phenomena that occur when intense few-cycle laser pulses interact with a solid, and opportunities related to this progress have recently begun to be explored.
In particular, implicit observations of subcycle temporal structures associated with interband tunneling in a dielectric were reported~\cite{Gertsvolf_2010_JPB_43_131002,Mitrofanov_2011_PRL_106_147401}, and effects related to Bloch oscillations in a bulk solid were observed~\cite{Ghimire_2011_NP_7_138,Ghimire_2011_PRL_107_167407}.

A direct observation of attosecond-scale electron motion was reported by Schiffrin \emph{et~al.} in a very recent paper~\cite{Schiffrin_2013_Nature_493_70} demonstrating that electric currents in a fused silica sample can be switched and driven by the instantaneous field of an optical waveform at intensities just below the damage threshold.
It was argued that, at field strengths where the induced potential difference between neighboring unit cells approaches the bandgap energy, the experiment can be interpreted using Wannier--Stark states.
In this paper, we show that these kinds of measurements can also be explained within a more conventional approach based on interference between different multiphoton channels.
This relates the results of Ref.~\onlinecite{Schiffrin_2013_Nature_493_70} to the scope of coherent control.

Injection and coherent control of electric currents in semiconductors was studied before, both in experiments and theory~\cite{Kurizki_1989_PRB_39_3435,Atanasov_1996_PRL_76_1703,%
Hache_1997_PRL_78_306,Rioux_2012_PE_45_1}.
Coherent control was demonstrated in ($\omega$, $2\omega$) schemes based on the interference of one- and two-photon excitation pathways induced by a laser pulse and its second harmonic.
In such experiments, it is sufficient for pulse durations and time delays to be less than or comparable to the carrier dephasing time ($\sim 100$~fs).
Now, with the availability of the few-cycle pulses, control over the photocurrent can be achieved \emph{within a single laser pulse} and on much shorter time scales ($\sim 1$~fs).

The interpretation of carrier-envelope phase effects in atomic and molecular systems in terms of quantum-mechanical interference is also well known~\cite{Nguyen-Dang_2005_PRA_71_023403,Nakajima_2006_PRL_96_213001}, and a general abstract theory that describes all phase effects in these terms is available~\cite{Roudnev_2007_PRL_99_220406}.
However, it is not obvious whether concepts developed for relatively weak fields can be applied to interpret experiments with intense few-cycle pulses, in which the absorption of many photons is required to excite valence-band electrons, and laser field amplitude reaches values at which perturbation theory is expected to break down.

To investigate the non-resonant optical injection and control of electronic currents in dielectrics, we develop a model based on a self-consistent solution of multiband optical Bloch equations (OBE)~\cite{Haug_2004,Rossi_2002_RMP_74_895} together with equations for the dielectric polarization and field inside the crystal.
The paper is organized as follows.
In Sec.~\ref{s:theory} we discuss our theoretical formalism, the gauge choice, and approximations that we make calculating the amount of charge flowing through a capacitor-like junction.
Results of our numerical simulations, their interpretation and comparison to experimental data are given in Sec.~\ref{s:num}.
Sec.~\ref{s:concl} presents our conclusions.

\section{Theory}\label{s:theory}

In order to calculate the current and polarization induced by an ultrashort pulse, we solve the density-matrix equations in the independent particle approximation.
Unlike the time-dependent Schr\"odinger equation~\cite{Bachau_2006_PRB_74_235215,Korbman_2013_NJP_15_013006}, this approach takes into account Pauli blocking of interband transitions~\cite{Axt_2004_RPP_67_433} and it can be extended to account for electron-electron scattering and interaction with the bath.
The approximation of independent particles is reasonable for ultrafast strong-field phenomena since the carrier-field interaction is much stronger than the carrier-carrier interaction.
Another argument in favor of this approximation is that the net current only depends on the total electronic momentum, which is not changed by electron-electron interaction~\cite{Datta_1995,Jacobini_2010}.
On timescales much shorter than a period of lattice oscillations, we can also neglect the electron-phonon interaction.
For longitudinal optical (LO) phonons in typical semiconductors and dielectrics, the oscillation period is about tens of femtoseconds.

In the basis of Bloch states, the Hamiltonian of the electronic subsystem of a dielectric interacting with a laser field can be written as
\begin{equation}\label{e:H}
  H = H_0 + H_{\mathrm{int}}(t),
\end{equation}
\begin{equation}\label{e:H0}
  H_0 = \sum_{\ell \in \text{CB}}
  \epsilon_{\ell}^{\mathrm{e}}
  c_{\ell}^\dagger c_{\ell}
+ \sum_{j \in \text{VB}}
  \epsilon_{j}^{\mathrm{h}}
  d_{j}^\dagger d_{j},
\end{equation}
\begin{equation}\label{e:Hint}
\begin{gathered}
  H_{\mathrm{int}}(t) = \sum_{\ell_1,\ell_2\in \text{CB}}
  \mathcal{M}_{\ell_1 \ell_2} (t)
  c_{\ell_1}^\dagger c_{\ell_2}
+ \sum_{j_1, j_2 \in \text{VB}}
  \mathcal{M}_{j_1 j_2} (t)
  d_{j_1}^\dagger d_{j_2}
\\
+ \sum_{\ell, j}
  \left[
    \mathcal{M}_{\ell j} (t)
    c_{\ell}^\dagger d_{j}^\dagger
  + \mathcal{M}_{\ell j}^* (t)
    d_{j} c_{\ell}
  \right],
\end{gathered}
\end{equation}
where $\ell = \{f, \vec k', s'\}$ and $j = \{i, -\vec k, s\}$ are composite indices that include a band index ($i$ and $f$), crystal momentum ($\vec k$ and $\vec k'$) and spin ($s$ and $s'$).

In the velocity gauge (VG), the amplitude of interaction with the optical field can be written as
\begin{multline}\label{e:MfiVG}
  \mathcal{M}_{fi}^{\mathrm{VG}}(\vec k', \vec k, t)
= - \sigma_{fi} \frac{e}{m_0} \vec A(t) \delta(\vec k'-\vec k)
  [ \delta_{fi}\hbar\vec k + \vec p_{fi}(\vec k) ]
\\
  + \delta(\vec k'-\vec k) \delta_{fi} \frac{e^2 A^2(t)}{2 m_0},
\end{multline}
where $\vec A(t)$ is the vector potential in a medium,
indices $i$ and $f$ enumerate all valence and conduction bands,
$\sigma_{fi}$ is a band-specific sign function defined as
\begin{equation}\label{e:bsgn}
  \sigma_{fi} = \left\{
  \begin{array}{ll}
    +,& i, f \in \mathrm{VB},\\
    -& \text{otherwise},\\
  \end{array}\right.
\end{equation}
and
\begin{equation}\label{e:pfi}
  \vec p_{fi}(\vec k) = -\frac{\mathrm{i}\hbar}{\Omega}
    \int_{\Omega} d^3 r\,
    u_{f,\vec k}^*(\vec r) \nabla_{\vec r} u_{i,\vec k}(\vec r)
\end{equation}
is the momentum matrix element between Bloch amplitudes $u_{i,\vec k}$,
$\Omega$ being the volume of an elementary cell.

The term proportional to $A^2(t)$, can be eliminated in the dipole approximation by the following gauge transformation~\cite{Bransden_2000}:
\begin{equation}
 \Psi'(t) = \Psi(t) \exp \left[
 \frac{\mathrm{i}}{\hbar} \frac{e^2}{2m_0} \int_{-\infty}^t A^2(t') dt'
 \right ].
\end{equation}

The expression for the interaction amplitude in the length gauge can be written as~\cite{Blount_1962,Aversa_1995_PRB_52_14636}
\begin{equation}\label{e:MfiLG}
 \mathcal{M}_{fi}^{\text{LG}} (\vec k', \vec k, t)
= -\sigma_{fi} e \vec E(t)
  [ \delta_{fi}\mathrm{i} \nabla_{\vec k} + \vec \xi_{fi}(\vec k) ] \delta(\vec k'-\vec k)
,\\
\end{equation}
where $\vec E(t)$ is the electric field in the medium, and
\begin{equation}\label{e:rfi}
  \vec\xi_{fi}(\vec k) = \frac{\mathrm{i}}{\Omega}
    \int_{\Omega} d^3 r\,
    u_{f,\vec k}^*(\vec r) \nabla_{\vec k} u_{i,\vec k}(\vec r)
.
\end{equation}

In the present paper, we use the velocity gauge because it results in a system of dynamic equations in which different $\vec k$-states are uncoupled.
In contrast, the corresponding length-gauge equations have a term proportional to $\nabla_{\vec k} \delta(\vec k'-\vec k)$, which couples different $\vec k$-points and introduces singularities, and matrix elements $\vec\xi_{fi}(\vec k)$, which are not uniquely defined.
These problems with the length gauge in the description of optical phenomena in solids are well-known, and have been solved in recent theoretical treatments~\cite{Aversa_1995_PRB_52_14636,Virk_2007_PRB_76_035213}.

Nevertheless, the treatment of the crystal polarization in the velocity gauge has a few important drawbacks, discussed, for example, in Ref.~\onlinecite{Virk_2007_PRB_76_035213}.
First, the solution of dynamic equations requires a large number of bands, only a few of which contain a significant carrier population at the end of a simulation.
Second, a certain sum rule for momentum matrix elements must be satisfied~\cite{Virk_2007_PRB_76_035213}.
Violation of these conditions leads to serious numerical artifacts and unphysical results.
In other words, a description of strong-field phenomena in the velocity gauge requires an accurate solution of the stationary problem.
Equations in the length gauge have an advantage of less strict requirements to the quality of the eigenproblem solution, so that the energy spectrum and optical matrix elements could be calculated with more simple and rough methods, or even used as adjustable parameters~\cite{Schiffrin_2013_Nature_493_70}.

Despite these problems, a fully \emph{ab~initio} approach based on time-dependent density functional theory in the velocity gauge was developed~\cite{Otobe_2008_PRB_77_165104,Yabana_2012_PRB_85_045134} and successfully applied to describe the non-perturbative polarization response of solids to a strong field.
Such calculations are very resource-demanding and require a high-performance supercomputer.

Our present approach is based on the numerical solution of density-matrix equations for a one-dimensional lattice and has very moderate computational requirements.
Yet it allows us to reproduce important features of laser-matter interaction on the ultrashort time scale for intensities near the damage threshold, and it gives good agreement with experimental data.

Starting from the definitions of two-point density-matrix elements,
\begin{gather*}
  \rho_{j\ell}^{(\vec k)} \equiv \langle d_{j,-\vec k} c_{\ell, \vec k} \rangle
,\quad
  \rho_{\ell \ell'}^{(\vec k)} \equiv \langle c_{\ell, \vec k}^\dagger c_{\ell', \vec k} \rangle
,\\
  \rho_{jj'}^{(\vec k)} \equiv \langle d_{j, -\vec k}^\dagger d_{j', -\vec k} \rangle,
\end{gather*}
Eqs.~\eqref{e:H}--\eqref{e:MfiVG}, and the Liouville--von~Neumann equation
\begin{equation*}
  \frac{d\rho}{d t} = \frac{1}{i\hbar} [H, \rho],
\end{equation*}
one obtains the following system of equations for each $\vec k$ in the Brillouin zone:
\begin{equation}\label{e:OBE}
\begin{aligned}
  \frac{d}{dt}\rho_{j\ell} = \frac{1}{i\hbar}\Biggl[
    &\sum_{j'} \Bigl(
      \mathcal{E}_{jj'} \rho_{j'\ell}
    - \mathcal{M}_{\ell j'} \rho_{j'j}
    \Bigr)
\\
  + &\sum_{\ell'} \Bigl(
      \mathcal{E}_{\ell\ell'} \rho_{j\ell'}
    - \mathcal{M}_{\ell'j } \rho_{\ell'\ell}
    \Bigr)
  + \mathcal{M}_{\ell j}
  \Biggr]
,\\
  \frac{d}{dt}\rho_{\ell\ell'} = \frac{1}{i\hbar} \Biggl[
  &\sum_{\ell''} \Bigl(
    \mathcal{E}_{\ell' \ell''} \rho_{\ell  \ell''}
  - \mathcal{E}_{\ell''\ell  } \rho_{\ell''\ell' }
  \Bigr)
\\
  + &\sum_j
  \Bigl(
    \mathcal{M}_{\ell' j} \rho_{j\ell}^*
  - \mathcal{M}_{\ell  j}^* \rho_{j\ell'}
  \Bigr)\Biggr]
,\\
  \frac{d}{dt}\rho_{jj'} = \frac{1}{i\hbar} \Biggl[
  &\sum_{j''} \Bigl(
    \mathcal{E}_{j' j''} \rho_{j  j''}
  - \mathcal{E}_{j''j  } \rho_{j''j' }
  \Bigr)
\\
+ &\sum_\ell \Bigl(
    \mathcal{M}_{\ell j'} \rho_{j\ell}^*
  - \mathcal{M}_{\ell j}^* \rho_{j' \ell}
  \Bigr)\Biggr]
,
\end{aligned}
\end{equation}
where
\begin{equation*}
  \mathcal{E}_{\ell\ell'}
= \delta_{\ell\ell'} \epsilon_\ell
+ \mathcal{M}_{\ell\ell'}
,
\end{equation*}
and indices $\ell$ and $j$ enumerate conduction and valence bands, respectively.
Quasimomentum indices are omitted here, for simplicity.

The charge carriers generated by an intense laser pulse induce a strongly nonlinear response.
If the characteristic size of the sample in the direction of polarization is small enough, then the field inside the crystal is affected by the surface charge distribution, which should be taken into account by solution of semiconductor Maxwell--Bloch equations~\cite{Hess_1996_PRA_54_3347,Giessen_1998_PRL_81_4260,Haug_2004}.
However, a full numerical solution of these equations in a three-dimensional case presents a rather hard computational problem, so we resort to a simplified model of the dielectric polarization.

\begin{figure}[!ht]
  \includegraphics[width=5cm]{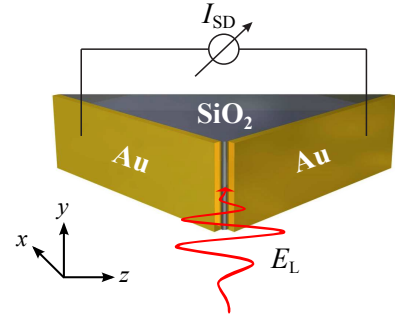}
  \caption{\label{f:exp_setup}%
    (Color online) A schematic representation of the metal-dielectric junction exposed to the electric field $E_{\mathrm{L}}$ of the laser pulse.
    Courtesy of A.~Schiffrin.
  }
\end{figure}

In the experiment reported in Ref.~\onlinecite{Schiffrin_2013_Nature_493_70}, the fused silica sample was exposed to a few-cycle pulse with a central photon energy~$\hbar\omega_{\mathrm{L}} \approx 1.7$~eV, and charge displaced by the optical field was collected by two gold electrodes (see Fig.~\ref{f:exp_setup}).

The dielectric slab is much thinner than the beam size ($\sim 100~\mu$m) but much thicker than the electron's de~Broglie wavelength.
In this case, the effects of quantum confinement are negligible, and thus the stationary states are described by Bloch wavefunctions.
An external field polarizes the sample, and the induced surface charge creates a screening field.
From the condition of continuity of the normal component of the electric displacement field at the surface between two media $\vec D_{1,\perp} = \vec D_{2,\perp}$ it follows that the electric field $\vec E(t)$ inside the sample is connected with the laser field $\vec E_{\mathrm{L}}(t)$ and macroscopic polarization $\vec P(t)$ by the following relation (SI units):
\begin{equation}\label{e:EELP}
  \vec E(t) = \vec E_{\mathrm{L}}(t) - \vec P(t)/\varepsilon_0.
\end{equation}
From the definition $\dot{\vec A} = -\vec E$, we obtain the equation for the vector potential inside the dielectric:
\begin{equation}\label{e:dotA}
  \frac{d}{dt}\vec A(t) = -\vec E_{\mathrm{L}}(t) + \vec P(t)/\varepsilon_0.
\end{equation}
The macroscopic polarization satisfies the equation
\begin{equation}\label{e:dotP}
  \frac{d}{dt}\vec P(t) = \vec J(t),
\end{equation}
where the charge current density is given by the following expression~\cite{Rammer_1998}:
\begin{multline}\label{e:Jdef}
  \vec J(t) = -\frac{e}{m_0} \frac{1}{V}\int_V d^3 r \sum_{s}
  \left[
    \frac{\hbar}{2 i}(\nabla_{\vec r} - \nabla_{\vec r'})
  + e \vec A(t)
  \right]
\\
  \times\rho(\vec r', s', \vec r, s; t) \biggr|_{\vec r' = \vec r, s' = s}
\end{multline}
where averaging is done over a volume $V$,
$s$ and $s'$ are spin indices of initial and final states, respectively, and
$m_0$ is the free-electron mass.

From this equation we obtain
\begin{multline}\label{e:J}
  \vec J(t) = \frac{2e}{m_0} \int_{\text{BZ}} \frac{d^3 k}{(2\pi)^3}
  \Biggl\{
    \sum_{f,i} \sigma_{fi} \rho_{fi}(\vec k, t)
\\
    \times\left[ \delta_{fi} \hbar \vec k + \Re \vec p_{fi}(\vec k) \right]
  - N_{\mathrm{VB}} e \vec A(t)
  \Biggr\},
\end{multline}
where indices $f$ and $i$ enumerate all valence and conduction bands,
the sign function $\sigma_{fi}$ is defined by Eq.~\eqref{e:bsgn},
the multiplier 2 accounts for the spin degeneracy,
and $N_{\mathrm{VB}}$ is the number of valence bands.

We solve Eqs.~\eqref{e:OBE},~\eqref{e:dotA}, and~\eqref{e:dotP} self-consistently for a given laser field $\vec E_{\mathrm{L}}(t)$, energy spectrum $\epsilon_{i}(\vec k)$, and momentum matrix elements~$\vec p_{fi}(\vec k)$.

As long as the external field is not strong enough to generate a significant number of electron-hole pairs, the induced screening field can be evaluated from the linear polarization response: $\vec D = \varepsilon \vec E$.
In the approximation of an instantaneous and linear, we evaluate the relative permittivity $\varepsilon$ at a central laser frequency $\omega_{\mathrm{L}}$ and, from Eq.~\eqref{e:EELP}, we obtain
\begin{equation}\label{e:AALeps}
  \vec A(t) = \vec A_{\mathrm{L}}(t)/\varepsilon.
\end{equation}
In this case only Eqs.~\eqref{e:OBE} for the density matrix must be solved.

In the next sections, we discuss the comparison of these two approaches, and we refer to the model that includes dielectric screening self-consistently [Eqs.~\eqref{e:OBE},~\eqref{e:dotA}, and~\eqref{e:dotP}] as OBE/SCDS, and to the linear screening model [Eqs.~\eqref{e:OBE} and~\eqref{e:AALeps}] just as OBE.

\section{Results and discussion}\label{s:num}

It is well known that nonparabolicity of electronic bands plays an important role in strong-field phenomena~\cite{Ghimire_2011_PRL_107_167407}, and their description requires an accurate solution of the stationary problem for the entire Brillouin zone.
We calculate the stationary electronic levels $\epsilon_{i,\vec k}$ and momentum matrix elements $\vec p_{fi}(\vec k)$ using the plane-wave pseudopotential method~\cite{Cohen_1988,Bachau_2006_PRB_74_235215}.

The eigenvalue problem for the crystal is written as
\begin{equation*}
  \left[ -\frac{\hbar^2}{2m_0} \nabla^2 + U(\vec r) \right] \phi_{n,\vec k}(\vec r)
= \epsilon_{n,\vec k} \phi_{n,\vec k}(\vec r),
\end{equation*}
where $\phi_{n,\vec k}(\vec r) = u_{n,\vec k}(\vec r) \exp(\mathrm{i} \vec k \vec r) $ is the wave function of an electron, $u_{n,\vec k}(\vec r)$ is a Bloch amplitude,
and $U(\vec r)$ is a periodical lattice potential.
Since $u_{n,\vec k}(\vec r)$ has the periodicity of the lattice, it can be expanded in a Fourier sum:
\begin{equation}
  u_{n,\vec k}(\vec r) = \sum_m C_{nm}^{(\vec k)} \exp(\mathrm{i} \vec K_m \vec r),
\end{equation}
where $\vec K_m$ are the reciprocal-lattice vectors.

Assuming that the laser pulse is linearly polarized, we may reduce the problem to one spatial dimension and solve the density-matrix equations for the $\vec k$-states that belong to a certain direction in the Brillouin zone.
In this paper, we consider a one-dimensional lattice with the pseudopotential
\begin{equation}\label{e:U}
  U(z) = c_1 (1 - \tanh^2 c_2 z ).
\end{equation}

\begin{figure}[!ht]
  \includegraphics[height=3.8cm]{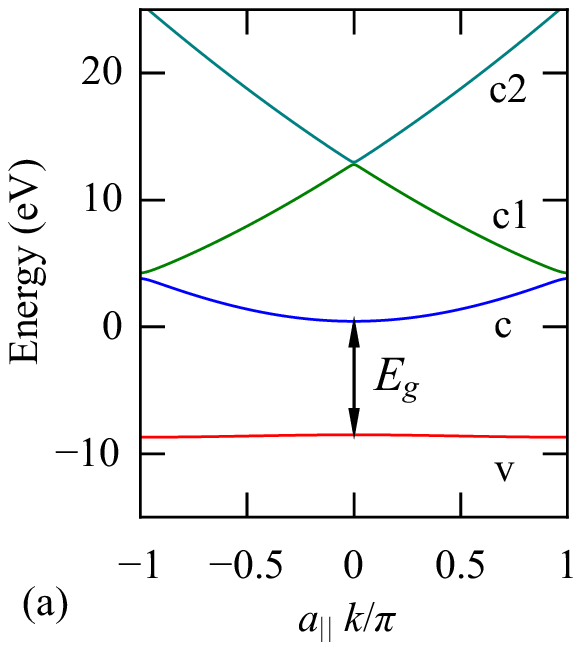}\quad
  \includegraphics[height=3.8cm]{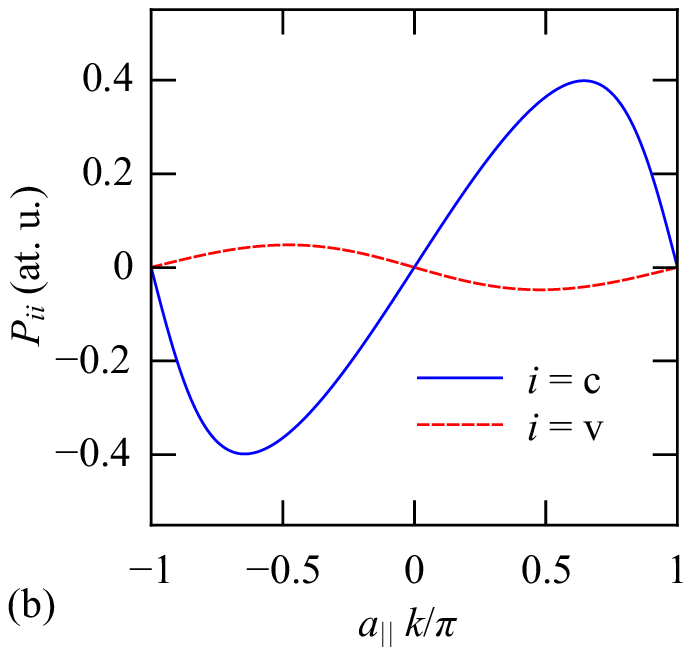}
  \caption{\label{f:k_E}%
    (Color online)
    (a) Band structure obtained via numerical solution of the eigenvalue problem for the one-dimensional periodic potential given by Eq.~\eqref{e:U}.
    (b) Diagonal matrix elements of the momentum operator $P_{ii} \equiv \hbar k + p_{ii}(k)$ versus $k$ for the topmost valence and lowest conduction bands.
  }
\end{figure}

We assume that the laser field polarization is parallel to the [001] direction of $\alpha$-quartz, for which the lattice constant is $a_{\parallel} \equiv c = 5.4$~\AA{}.
Fit parameters $c_1 = -2.2$ and $c_2 = 0.9$ were chosen such that the energy gap is~$E_g \approx 9$~eV and the effective mass of an electron in the first conduction band is~$m_{\mathrm{c}} \approx 0.4 m_0$ [see Fig.~\ref{f:k_E}(a)].
These values are in good agreement with the results for the $\Gamma$--A direction in the Brillouin zone of $\alpha$-quartz obtained with more complex and rigorous treatments~\cite{Schneider_1976_PRL_36_425,Chelikowsky_1977_PRB_15_4020,Gnani_2000_ITED_47_1795}.

It is worth noting that diagonal momentum matrix elements between the full Bloch functions $P_{ii} \equiv \hbar k + p_{ii}(k)$ quickly approach zero at the boundaries of the Brillouin zone, i.e. electrons are slowing down in this region [see Fig.~\ref{f:k_E}(b)].

To produce results that could be compared with experimental data, we must specify a proper connection between our one-dimensional calculations and physical values defined for a three-dimensional crystal.
We assume that the electromagnetic wave is polarized along the $z$-axis and the current density slowly changes within the $xy$-plane.
Then the current density $\vec J(t)$ in a three-dimensional (3D) crystal is connected with its 1D counterpart via the effective cross-section $\mathfrak{S}$ of the three-dimensional Brillouin zone,
\begin{equation*}
  \vec J(t)
= \int_{\mathrm{BZ}} \frac{d^3 k}{(2\pi)^3}\,\vec j(\vec k, t)
= \vec n \frac{\mathfrak{S}}{(2\pi)^2} \int_{-\pi/a_{\parallel}}^{\pi/a_{\parallel}} \frac{d k_z}{2\pi}\,j_{\mathrm{1D}}(k_z, t),
\end{equation*}
where $j_{\mathrm{1D}}(k_z, t)$ is the one-dimensional analog of the expression in curly brackets from Eq.~\eqref{e:J}, and $\vec n$ is the unit vector in the direction of the current.
In a first approximation, the parameter~$\mathfrak{S}$ can be estimated directly from the properties of the lattice for a given material.
For $\alpha$-quartz with a hexagonal lattice and current direction along the [001] direction, we get
\begin{equation*}
  \mathfrak{S}_{\text{SiO}_2} = \frac{3\sqrt{3}}{2} \left(\frac{2\pi}{a_{\perp}}\right)^2 \approx 1.2~\text{at.\,u.},
\end{equation*}
where $a_{\perp} \equiv a = 4.9$~\AA{} is the lattice parameter of the crystal plane perpendicular to the direction of the laser field polarization.

On the other hand, the value of $\mathfrak{S}$ can be found more precisely from the requirement that the OBE solution should give a correct linear response in a low-intensity region.
With a given linear susceptibility $\chi^{(1)}$ and polarization calculated from Eq.~\eqref{e:OBE}, we have $\mathfrak{S}_{\text{SiO}_2} \approx 1.43$~at.\,u., which is in good agreement with the above estimation.

For numerical integration of Eqs.~\eqref{e:OBE},~\eqref{e:dotA}, and \eqref{e:dotP}, we discretized the $\vec k$-space and applied the fourth-order Runge--Kutta method with a constant stepsize.
At the highest field intensities considered in this paper, numerical convergence was achieved with 18 bands and 201 $k$-points.
The requirement of a large number of bands is a consequence of the velocity gauge drawbacks discussed earlier in Sec.~\ref{s:theory}.

In subsequent calculations, we use the following expression for the vector potential of the laser field:
\begin{gather}
  \vec A_{\mathrm{L}}(t) = -\vec A_0 \theta(\tau_{\mathrm{L}} - |t|)
  \cos^4\left( \dfrac{\pi t}{2 \tau_{\mathrm{L}}} \right)
  \sin( \omega_{\mathrm{L}} t + \varphi_{\mathrm{CE}} ),
\end{gather}
where $\omega_{\mathrm{L}}$ is the central frequency of the laser pulse,
$\theta(x)$ is the Heaviside step function,
$\vec A_0 = \vec E_0/\omega_{\mathrm{L}}$ and $\vec E_0$ are the amplitudes of vector potential and electric field in the vacuum, respectively,
$\varphi_{\mathrm{CE}}$ is the carrier-envelope phase.
Total pulse duration $\tau_{\mathrm{L}}$ is related to FWHM of $|\vec A(t)|^2$ as
$\tau_{\mathrm{L}} = \pi \tau_{\mathrm{FWHM}}/[4\arccos(2^{-1/8})] \approx 1.9\,\tau_{\mathrm{FWHM}}$.

\begin{figure*}[!ht]
  \includegraphics[width=8cm]{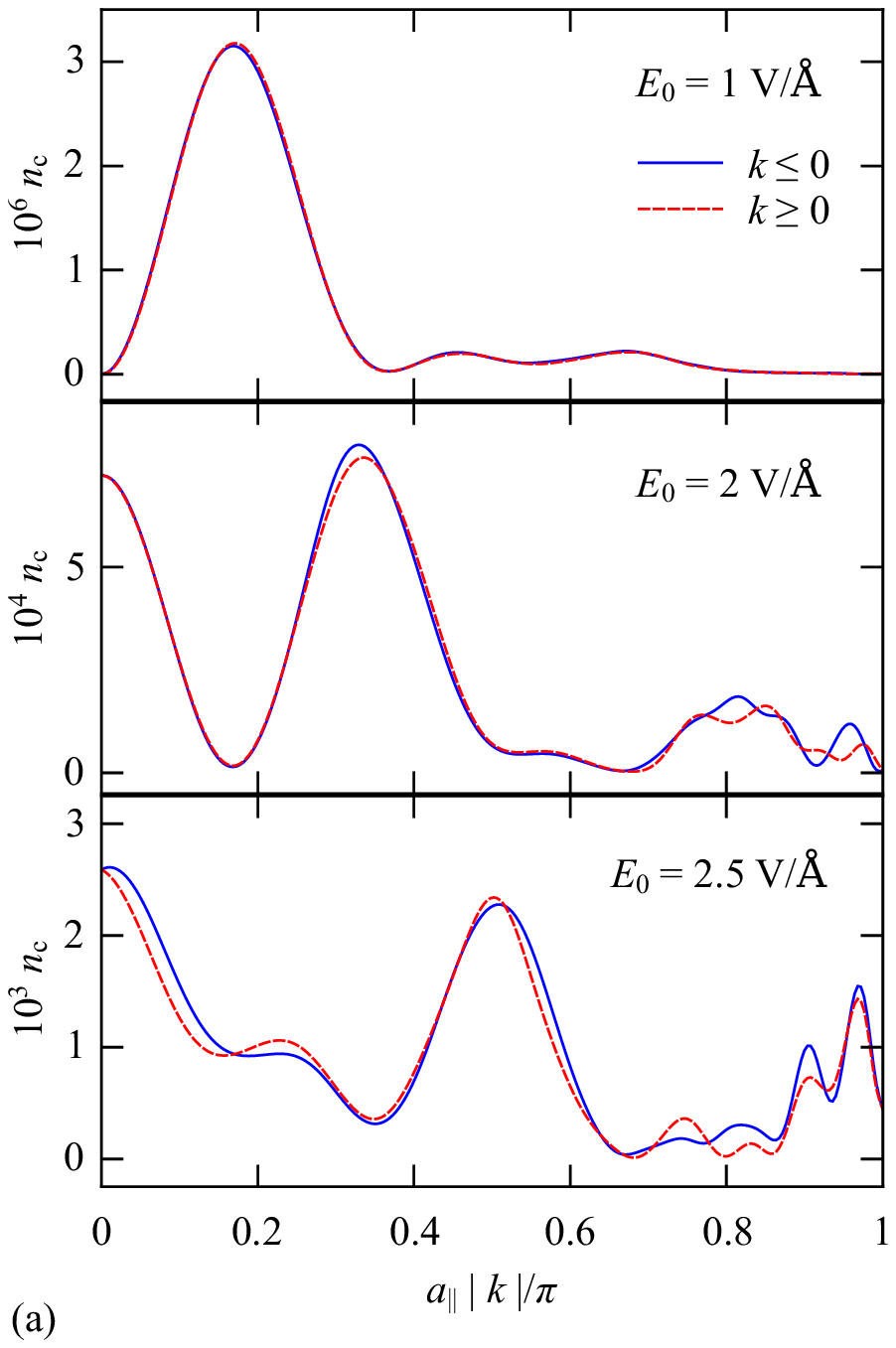}
  \includegraphics[width=8cm]{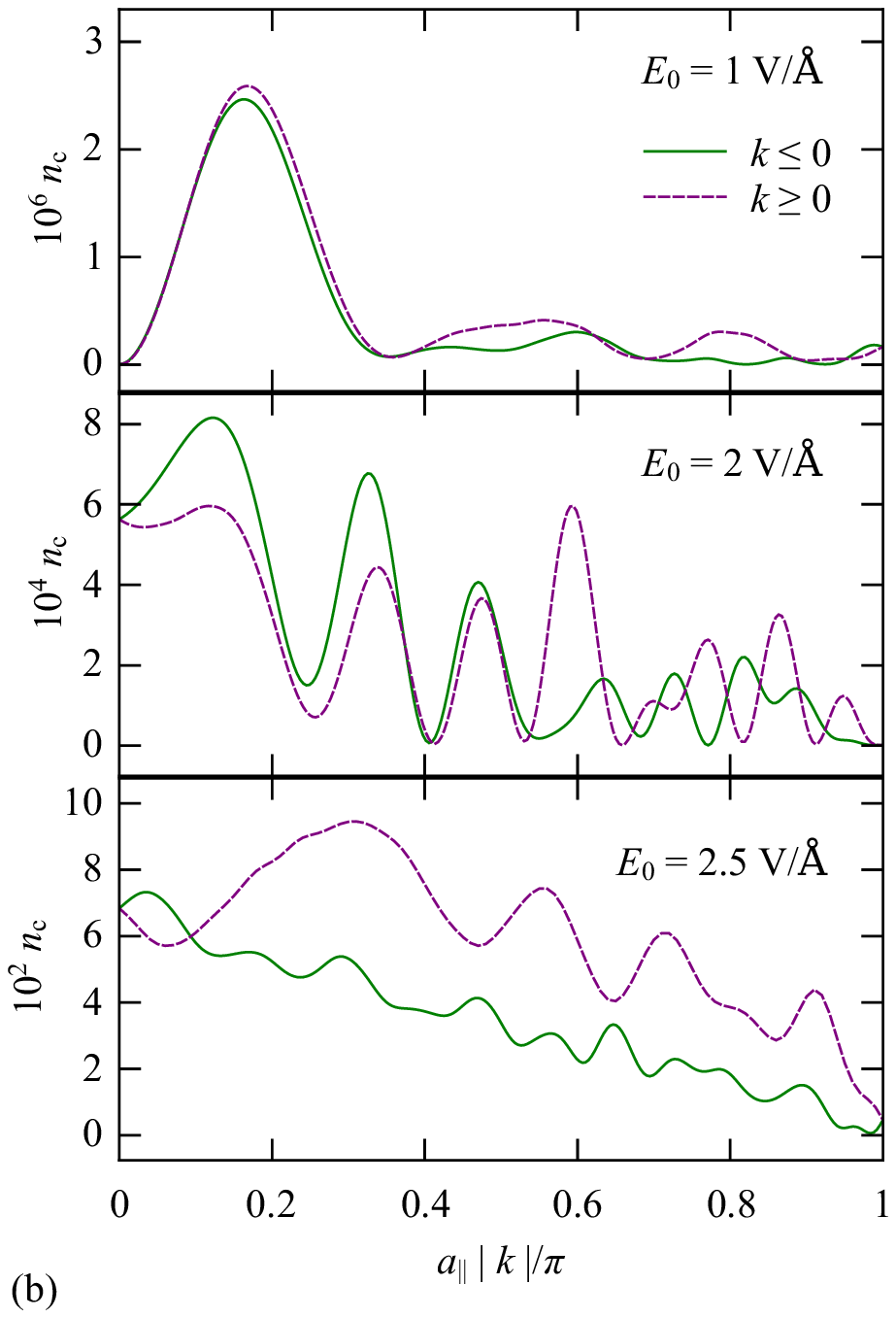}
  \caption{\label{f:k_nc}%
    (Color online) Distributions of the lowest conduction band population $n_{\text{c}}$ after the laser pulse ($\lambda_{\mathrm{L}} = 800$~nm, $\mathrm{FWHM} = 4$~fs, $\phi_{\mathrm{CE}} = 0$) for different field amplitudes, calculated (a) with OBE and (b) with OBE/SCDS.
    Populations for positive and negative crystal momenta are shown with the dashed and solid lines, respectively.
  }
\end{figure*}

\begin{figure}[!ht]
  \includegraphics[width=5.5cm]{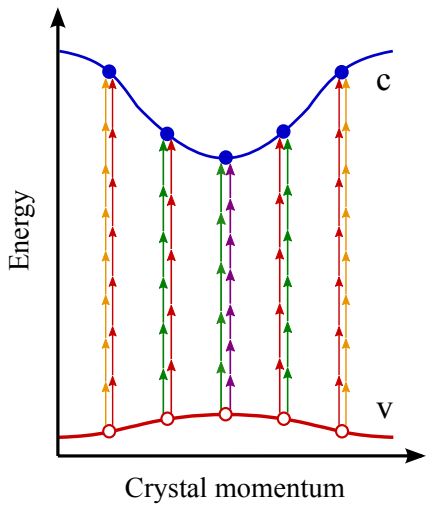}
  \caption{\label{f:bands}%
    (Color online) A schematic representation of multiphoton channel interference at different points in the Brillouin zone.
  }
\end{figure}

Let us start the discussion of our numerical results with the distribution of populations in the first conduction band at the moment of time when the electromagnetic field is over.
Fig.~\ref{f:k_nc} shows that the few-cycle pulse generates an asymmetric band population distribution in the crystal momentum space, which causes the appearance of macroscopic current.

We argue that these symmetry-breaking effects of the waveform with a well-defined CE phase emerge due to the interference of different multiphoton excitation pathways.
This phenomenon is schematically depicted on Fig.~\ref{f:bands}, where arrows of different colors describe different multiphoton channels that interfere with each other.
Note, that excitation pathways might have the same number of photons with different frequencies, as well as different numbers of photons.
In this picture, physical observables become CEP-dependent if the pulse is short enough, i.\,e.\ when its spectral width allows for existence of multiphoton channels with odd and even numbers of photons for the same $\vec k$-point in the Brillouin zone.
The interference of optical excitation pathways~\cite{Fortier_2004_PRL_92_147403} might be constructive for $\vec k$ and destructive for $-\vec k$.
When the laser pulse is over, electron-phonon collisions become the dominant interaction, which quickly restores the symmetry of electronic population distribution and make the current disappear.

Figs.~\ref{f:k_nc}(a) and (b) show that the maximum population distribution is shifted from the center, and the electron excitations spread over the whole Brillouin zone when the field intensity is increased.
This situation is strongly contrary to the case of a weak field, where most of the charge carriers are situated around the extremal points in the Brillouin zone.
At a relatively low field amplitude of $E_0 \sim 1$~V/\AA{}, both OBE and OBE/SCDS models give very similar results, as expected, but for stronger fields, the second one shows the overall increase of the population and even more asymmetric distribution.
This means that the screening field becomes sufficiently strong to cause interband transitions.

\begin{figure}[!ht]
  \includegraphics[width=7.5cm]{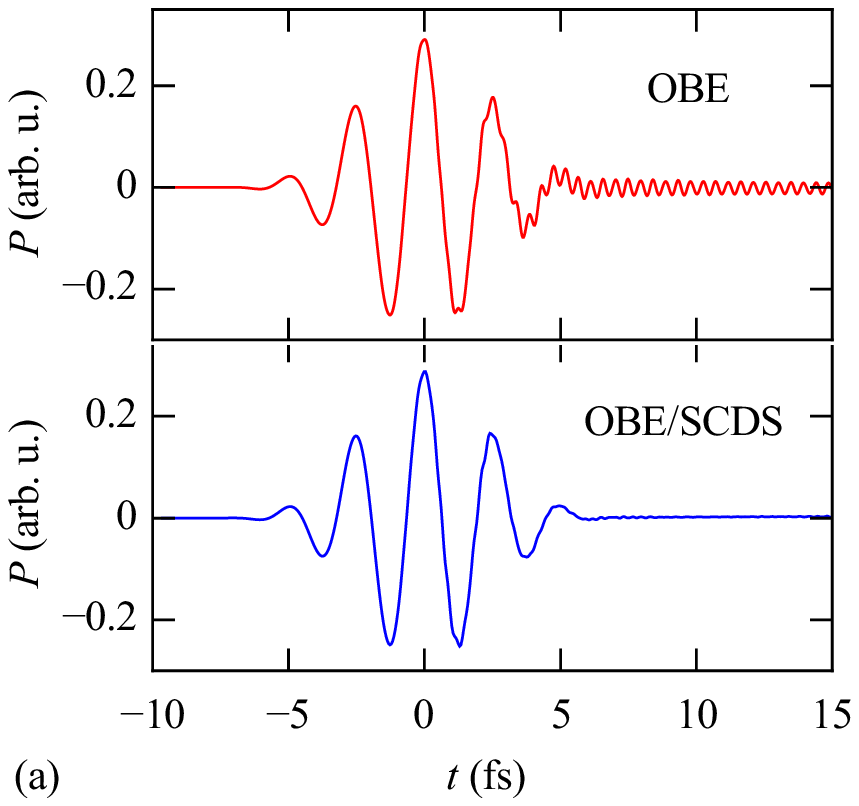}
  \includegraphics[width=8cm]{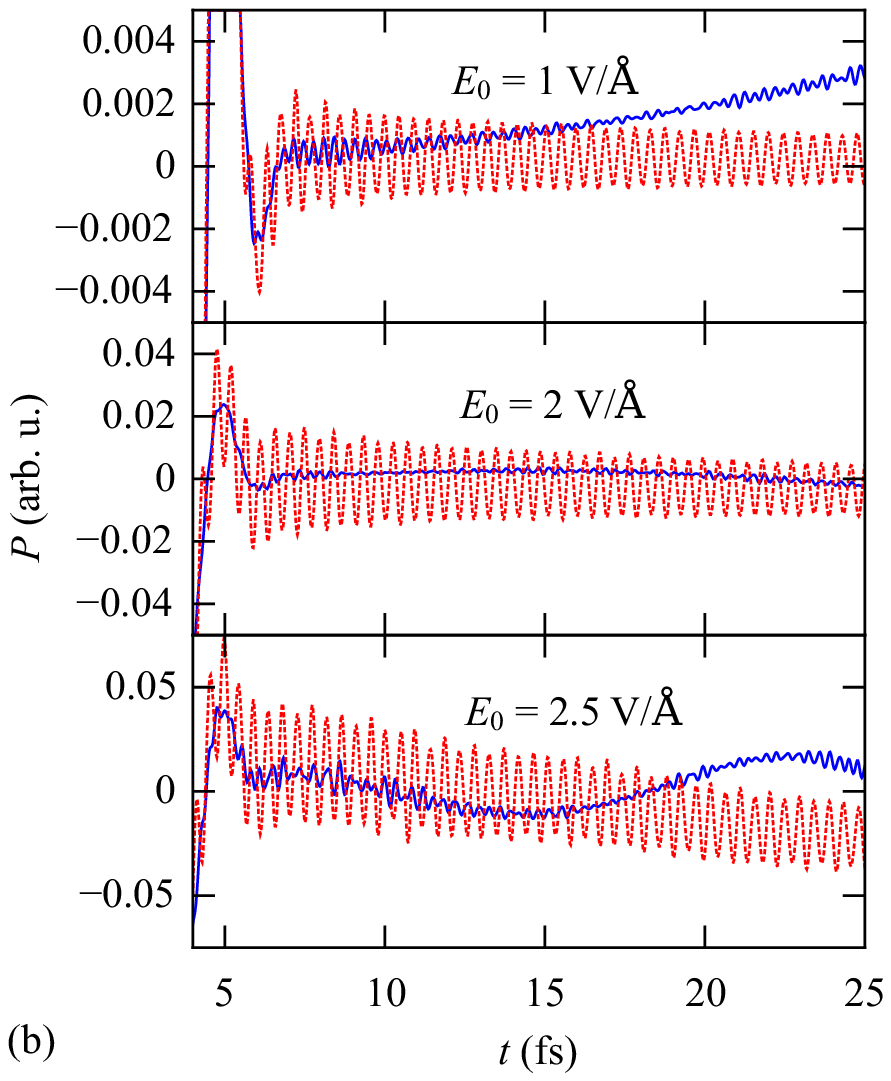}
  \caption{\label{f:t_P}%
    (Color online) (a) Polarization versus time calculated with OBE (top panel) and with OBE/SCDS (bottom panel), $E_0 = 2$~V/\AA{}, $\lambda_{\mathrm{L}} = 800$~nm, $\varphi_{\text{CE}} = 0$.
    (b) Polarization versus time for OBE (dotted line) and OBE/SCDS (solid line) at the end of the laser pulse.
    For the field intensities close to the damage threshold, the low-frequency oscillations in polarization response are observed.
  }
\end{figure}

For a deeper understanding of these observations, let us consider the polarization response.
Fig.~\ref{f:t_P}(a) depicts a typical time-dependent polarization in response to a laser pulse with a peak intensity near the damage threshold ($E_0 \sim 2$~V/\AA{}).
The polarization response within the first half of the pulse is linear because the field is not strong enough to excite charge carriers to the conduction band, so the contribution to the polarization comes only from small displacements of electronic shells that form the valence bands of the crystal.
Once the field reaches a certain strength, electron-hole pairs are created, and the polarization response shows quantum beats with a characteristic energy approximately equal to the band gap.

A comparison of the plots in Fig.~\ref{f:t_P}(a) shows that, with the self-consistently evaluated screening field which remains after the laser pulse, quantum beats are significantly reduced.
The suppression of quantum beats in a strong field was already pointed out in our previous  paper~\cite{Korbman_2013_NJP_15_013006}.
In addition to that, the interaction of electrons with the screening field induced by quantum beats results in a phenomenon analogous to stimulated emission.
The induced screening field $\vec E(t)$ oscillates out of phase with respect to the polarization $\vec P(t)$.
It drives transitions from the conduction band to the valence band, and thus, decreases the amplitude of quantum beats.

The time dependence of polarization [Fig.~\ref{f:t_P}(b)] shows that the self-consistent screening model predicts low-frequency oscillations that persist after the laser pulse.
These oscillations occur as a result of collective electron motion driven by the surface charge field.
For the parameters of our simulations, the frequency of these oscillations is in the terahertz region (e.\,g. $50$~THz for $E_0 = 2.5$~V/\AA{}), which is close to the frequency of plasmonic oscillations, although, in our model, we account for electron-electron interaction only implicitly, via the macroscopic screening field.
In principle, these oscillations should produce the electromagnetic radiation in the corresponding frequency range.
The phenomenon of THz emission from cold plasma oscillations in semiconductors excited by femtosecond optical pulses is well known and has been reported in a number of recent papers~\cite{Kersting_1997_PRL_79_3038,Bonitz_2000_PRB_62_15724,Meinert_2000_PRB_62_5003}.
Also, there is an important analogy with the optical rectification effect~\cite{Bass_1962_PRL_9_446,Rice_1994_APL_64_1324}.

Since we do not consider relaxation phenomena (radiative or non-radiative), these plasma oscillations do not decay with time in our simulations.
Their amplitude might even grow, if the screening field is large enough to induce interband transitions.
Also, the dielectric polarization model does not take into account the energy loss due to the emission of electromagnetic radiation, so that the dielectric acts like a resonator for the induced currents.
Consequently, the results of our present simulations are valid only within a small interval of a few femtoseconds after the laser pulse.
A more accurate description of the strong-field phenomena requires the complete solution of semiconductor Maxwell--Bloch equations that take into account electron-electron and electron-phonon interactions, as well as electromagnetic radiation from accelerated charge carriers.

In the experiment reported by Schiffrin~\emph{et~al.} in Ref.~\onlinecite{Schiffrin_2013_Nature_493_70}, the measurements yielded the total transferred charge $Q_{\mathrm{P}}$ in a fused silica junction defined as
\begin{equation}\label{e:QP}
  Q_{\mathrm{P}} = A_{\mathrm{eff}} q_{\mathrm{P}},
\end{equation}
where
\begin{equation}\label{e:qP}
  q_{\mathrm{P}} = \vec P(t \rightarrow \infty) \equiv \int_{-\infty}^{\infty} \vec J(t) dt
\end{equation}
is the transferred charge density, and $A_{\mathrm{eff}}$ is the effective cross section of the active volume, estimated as $\sim 5\times 10^{-12}$~m$^2$.

This simple definition is directly applicable if theory includes all necessary relaxation mechanisms, both radiative and non-radiative, and the integral of charge current density over time takes a finite value.
In order to compare our numerical results with the experiment, we need to estimate the value of integral~\eqref{e:qP}, taking into account the applicability limitations due to the absence of relaxation mechanisms in our OBE/SCDS model.
Assuming that the current quickly decays after the laser pulse, we estimate the integral~\eqref{e:qP} by the value of polarization right after the pulse.
Thus, we can define the transferred charge density as an average value of the polarization taken in a time interval that is larger than the period of quantum beats $T_{\mathrm{b}}$ and smaller than the period of the long-wave polarization oscillations $T_{\mathrm{P}}$ after the laser pulse at the highest considered field intensity,
\begin{equation*}
  q_{\mathrm{P}} = \frac{1}{\Delta t} \int_{\tau_{\mathrm{L}}}^{\tau_{\mathrm{L}} + \Delta t} P(t) dt
,\quad
 T_{\mathrm{b}} \ll \Delta t \ll T_{\mathrm{P}}
.
\end{equation*}
In the subsequent calculations, we assume that $\Delta t = 1$~fs.

To convert the calculated values from atomic to SI units, we use the following relation:
\begin{equation*}
  q_{\mathrm{P}}\,\mathrm{(C/m^2)} = \frac{e}{a_{\mathrm{B}}^2}
  \frac{\mathfrak{S}}{(2\pi)^2} \times q_{\mathrm{P}}\,\mathrm{(at.\,u.)}
\approx 2.077 \times q_{\mathrm{P}}\,\mathrm{(at.\,u.)},
\end{equation*}
where $a_{\mathrm{B}}$ is the Bohr radius.

\begin{figure}[!ht]
  \includegraphics[width=7cm]{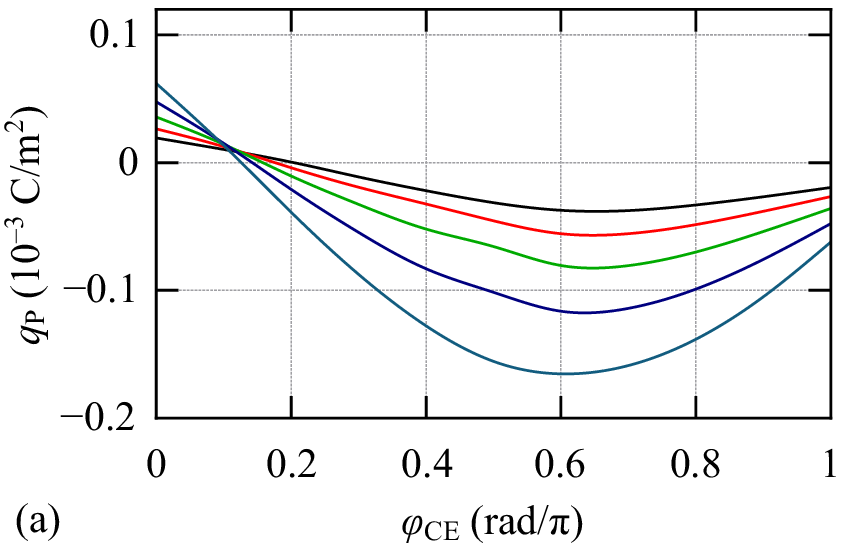}\quad
  \includegraphics[width=7cm]{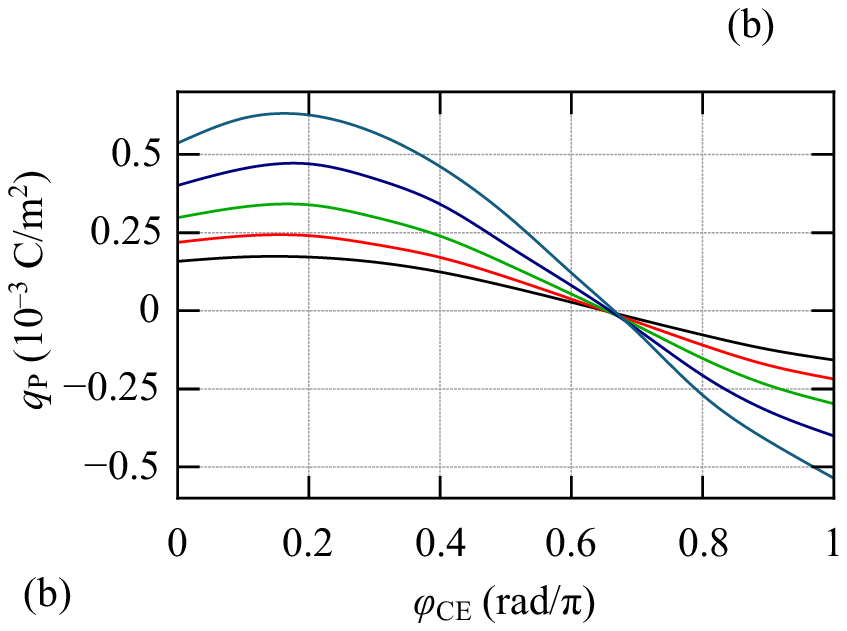}
  \caption{\label{f:CEP_qP}%
    (Color online) Transferred charge density $q_{\mathrm{P}}$ versus carrier-envelope phase obtained from
    (a) OBE and (b) OBE/SCDS calculations with the following values of the field amplitude: $1.15$, $1.2$, $1.25$, $1.3$, and $1.35$ V/\AA{}.
  }
\end{figure}

Fig.~\ref{f:CEP_qP} shows typical dependencies of the transferred charge density on the absolute value of the carrier-envelope phase.
The most striking difference between the results obtained with OBE (left panel) and those with OBE/SCDS (right panel) is the large shift (about $0.4 \pi$) of the CE phase that maximizes the transferred charge.
This phase shift has a simple explanation: in the model with constant screening the phase of the field inside the crystal is exactly the same as that of the laser field.
In the self-consistent screening model, the field inside the medium receives an additional phase shift from the polarization generated by the macroscopic charge current and quantum beats.

\begin{figure}[!ht]
  \includegraphics[width=6.5cm]{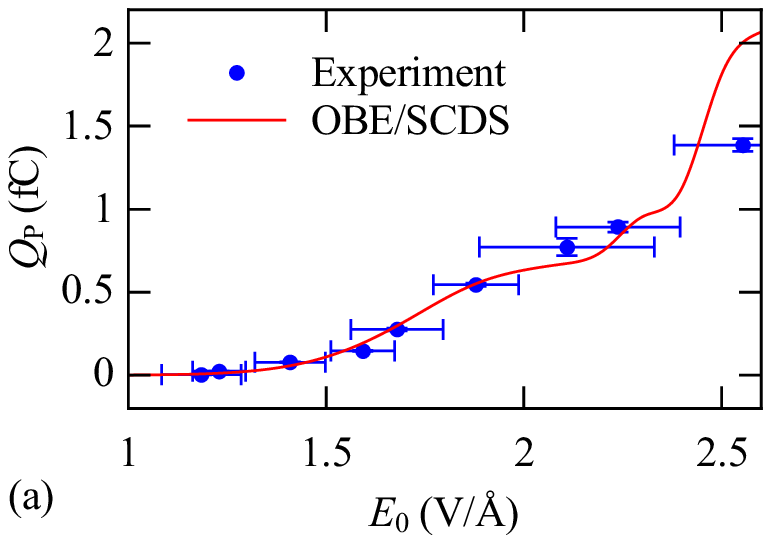}
  \includegraphics[width=7cm]{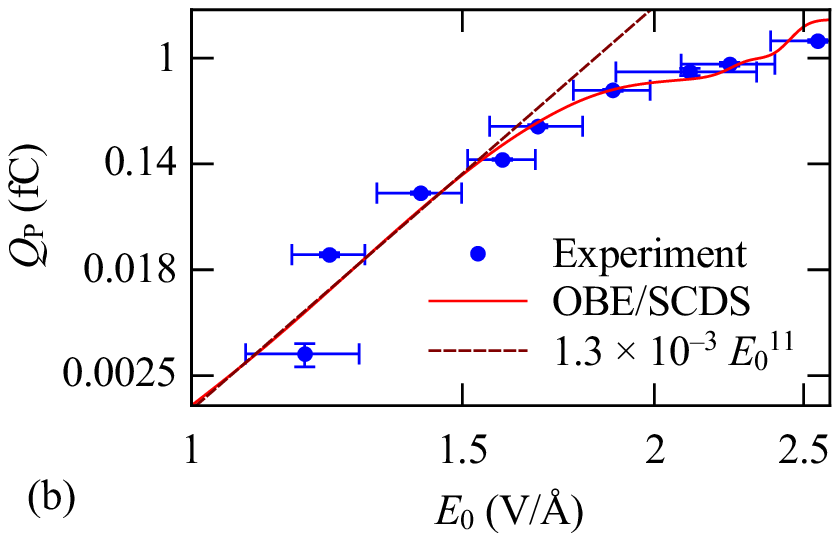}
  \caption{\label{f:E_QP}%
    (Color online) (a) Intensity scan of the transferred charge maximized over CEP: comparison of OBE/SDCS calculations with experimental data.
    (b) Results of fitting the calculated transferred charge dependence (solid line) and experimental data on a double-logarithmic plot with the power function of field amplitude.
  }
\end{figure}

As a justification of our model, we provide a comparison of our results with the measurements published in Ref.~\onlinecite{Schiffrin_2013_Nature_493_70}.
Since a self-consistent evaluation of the screening field is essential for very strong fields, we discuss only the OBE/SCDS model.
Fig.~\ref{f:E_QP}(a) shows very good agreement between theory and experiment up to $E_0 = 2.2$~V/\AA{}.
The discrepancy at high fields probably appears because the independent-particle approximation and dielectric screening model lose their applicability.

At low field intensities, the major contribution to the transferred charge should come from the interference between 5- and 6-photon channels.
In this case, $Q_{\mathrm{P}}$ is expected to scale as $E_0^{5+6}$, and we indeed find that the function $\propto E_0^{11}$ gives a good fit for $E_0 \lesssim 1.5$~V/\AA{} [Fig.~\ref{f:E_QP}(b)].
For more intense pulses, perturbation theory breaks down, and the transferred charge cannot be fitted by a simple power function.

\begin{figure}[!ht]
  \includegraphics[width=6.5cm]{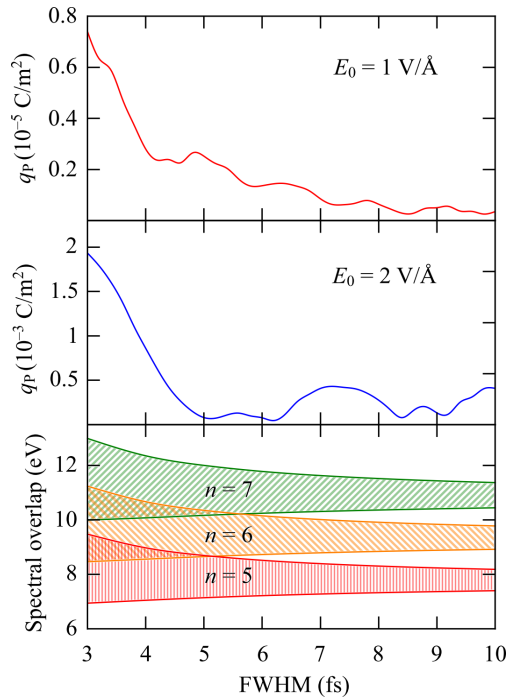}
  \caption{\label{f:FWHM_qP}%
    (Color online)
    Top panel: CEP-optimized transferred charge density versus FWHM of the laser pulse for high- and low-field amplitudes.
    Bottom panel: spectral overlap for 5-, 6-, and 7-photon excitation probabilities as a function of FWHM.
  }
\end{figure}

To verify the interpretation of the CEP-dependent currents as a consequence of the interference of multiphoton channels, we investigate the charge density as a function of pulse duration for a fixed central frequency and peak intensity of the laser pulse [Fig.~\ref{f:FWHM_qP}].
In the lower panel of Fig.~\ref{f:FWHM_qP}, we depict the energies accessible via different multiphoton channels.
The shaded areas in this figure show the regions where the probability distribution related to $n$-photon absorption
\begin{equation*}
  p_n(\omega) \sim \mathcal{F}[I^{n}(t)]
\end{equation*}
exceeds 90\% of its peak value.
Here $I(t)$ is the cycle-averaged intensity.

Even though a longer pulse creates more charge carriers, the transferred charge decreases with the pulse duration for both low and high field amplitudes (Figs.~\ref{f:FWHM_qP}(a) and (b), respectively).
This can be explained by the decreasing overlap of multiphoton channels (see the bottom panel in Fig.~\ref{f:FWHM_qP}).

\begin{figure}[!ht]
  \includegraphics[width=7cm]{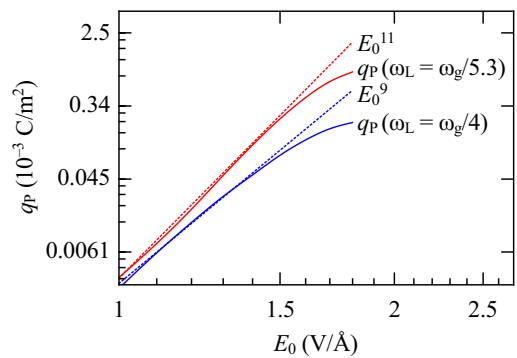}
  \caption{\label{f:E_lambda_qP}%
    (Color online) Double-logarithmic plot of CEP-optimized transferred charge density and its fit with powers of $E_0$ versus field amplitude for different central frequencies of the laser pulse (FWHM = 4~fs).
  }
\end{figure}

Second, we consider the field amplitude scans of the transferred charge density for the pulses with different central frequencies.
Fig.~\ref{f:E_lambda_qP} shows that the numerically calculated transferred charge for the scans in the region 1--1.5~V/\AA{} may be fitted with power functions $E^{2n+1}$ that correspond to the interference between $n$ and $(n + 1)$ photon channels.
Despite the success of this interpretation, it is clear that this simple power law cannot be valid under all circumstances.
For example, interference between more than two multiphoton channels, transitions to higher conduction bands, and non-perturbative phenomena affect the scaling law.

\section{Conclusions}\label{s:concl}

We have presented a quantum-kinetic approach for the description of strong-field non-resonant injection of photocurrents in dielectrics and wide bandgap semiconductors.
In this model, the appearance of a non-zero transferred electric charge and its dependence on the carrier-envelope phase are intrinsically related to an asymmetry of electronic population distribution in $\vec k$-space.
This asymmetry, which also remains after the laser pulse, can be explained by the interference of different multiphoton excitation channels.
We have presented two arguments supporting this interpretation.
First, we varied the bandwidth of the laser pulse and found that the amount of transferred charge is largely determined by the overlap of the dominant excitations channels.
Second, we have shown that, for fields up to $E_0\sim 1.5$~V/\AA{}, the dependence of the transferred charge on the amplitude of the laser pulse is well described by the simple perturbative expression for interfering multiphoton excitation pathways.
The scaling law predicted by our numerical model is in very good agreement with the experimental data published in Ref.~\onlinecite{Schiffrin_2013_Nature_493_70}, except for the largest values of pulse amplitude $E_0$, where the concentration of excited charge carriers reaches a level where our mean-field description becomes inappropriate and plasma oscillations start to determine the polarization response.

We have also compared two models for the evaluation of the screening field due to charges appearing on surfaces of a mesoscopic structure exposed to a laser pulse: a model assuming an instantaneous linear dielectric response, and a more rigorous one where the polarization response is evaluated self-consistently with quantum dynamics.
We have found that the self-consistent evaluation of the screening field is essential for an accurate description of CEP effects.
For instance, these two models predict different dependencies of the transferred charge on the carrier-envelope phase and, thus, the different values of $\phi_{\mathrm{CE}}$ which maximize the current.

\section{Acknowledgments}\label{s:ack}

We gratefully acknowledge A.~Schiffrin, N.~Karpowicz, T.~Paasch-Colberg, and Prof. F.~Krausz for fruitful discussions.
This work is supported by the DFG Cluster of Excellence: Munich-Centre for Advanced Photonics.

\bibliographystyle{apsrev4-1}
\bibliography{article}

\begin{thebibliography}{38}%
\makeatletter
\providecommand \@ifxundefined [1]{%
 \@ifx{#1\undefined}
}%
\providecommand \@ifnum [1]{%
 \ifnum #1\expandafter \@firstoftwo
 \else \expandafter \@secondoftwo
 \fi
}%
\providecommand \@ifx [1]{%
 \ifx #1\expandafter \@firstoftwo
 \else \expandafter \@secondoftwo
 \fi
}%
\providecommand \natexlab [1]{#1}%
\providecommand \enquote  [1]{``#1''}%
\providecommand \bibnamefont  [1]{#1}%
\providecommand \bibfnamefont [1]{#1}%
\providecommand \citenamefont [1]{#1}%
\providecommand \href@noop [0]{\@secondoftwo}%
\providecommand \href [0]{\begingroup \@sanitize@url \@href}%
\providecommand \@href[1]{\@@startlink{#1}\@@href}%
\providecommand \@@href[1]{\endgroup#1\@@endlink}%
\providecommand \@sanitize@url [0]{\catcode `\\12\catcode `\$12\catcode
  `\&12\catcode `\#12\catcode `\^12\catcode `\_12\catcode `\%12\relax}%
\providecommand \@@startlink[1]{}%
\providecommand \@@endlink[0]{}%
\providecommand \url  [0]{\begingroup\@sanitize@url \@url }%
\providecommand \@url [1]{\endgroup\@href {#1}{\urlprefix }}%
\providecommand \urlprefix  [0]{URL }%
\providecommand \Eprint [0]{\href }%
\providecommand \doibase [0]{http://dx.doi.org/}%
\providecommand \selectlanguage [0]{\@gobble}%
\providecommand \bibinfo  [0]{\@secondoftwo}%
\providecommand \bibfield  [0]{\@secondoftwo}%
\providecommand \translation [1]{[#1]}%
\providecommand \BibitemOpen [0]{}%
\providecommand \bibitemStop [0]{}%
\providecommand \bibitemNoStop [0]{.\EOS\space}%
\providecommand \EOS [0]{\spacefactor3000\relax}%
\providecommand \BibitemShut  [1]{\csname bibitem#1\endcsname}%
\let\auto@bib@innerbib\@empty
\bibitem [{\citenamefont {Gertsvolf}\ \emph {et~al.}(2010)\citenamefont
  {Gertsvolf}, \citenamefont {Spanner}, \citenamefont {Rayner},\ and\
  \citenamefont {Corkum}}]{Gertsvolf_2010_JPB_43_131002}%
  \BibitemOpen
  \bibfield  {author} {\bibinfo {author} {\bibfnamefont {M.}~\bibnamefont
  {Gertsvolf}}, \bibinfo {author} {\bibfnamefont {M.}~\bibnamefont {Spanner}},
  \bibinfo {author} {\bibfnamefont {D.~M.}\ \bibnamefont {Rayner}}, \ and\
  \bibinfo {author} {\bibfnamefont {P.~B.}\ \bibnamefont {Corkum}},\ }\href
  {\doibase 10.1088/0953-4075/43/13/131002} {\bibfield  {journal} {\bibinfo
  {journal} {J. Phys. B: At. Mol. Opt. Phys.}\ }\textbf {\bibinfo {volume}
  {43}},\ \bibinfo {pages} {131002} (\bibinfo {year} {2010})}\BibitemShut
  {NoStop}%
\bibitem [{\citenamefont {Mitrofanov}\ \emph {et~al.}(2011)\citenamefont
  {Mitrofanov}, \citenamefont {Verhoef}, \citenamefont {Serebryannikov},
  \citenamefont {Lumeau}, \citenamefont {Glebov}, \citenamefont {Zheltikov},\
  and\ \citenamefont {Baltu\ifmmode~\check{s}\else
  \v{s}\fi{}ka}}]{Mitrofanov_2011_PRL_106_147401}%
  \BibitemOpen
  \bibfield  {author} {\bibinfo {author} {\bibfnamefont {A.~V.}\ \bibnamefont
  {Mitrofanov}}, \bibinfo {author} {\bibfnamefont {A.~J.}\ \bibnamefont
  {Verhoef}}, \bibinfo {author} {\bibfnamefont {E.~E.}\ \bibnamefont
  {Serebryannikov}}, \bibinfo {author} {\bibfnamefont {J.}~\bibnamefont
  {Lumeau}}, \bibinfo {author} {\bibfnamefont {L.}~\bibnamefont {Glebov}},
  \bibinfo {author} {\bibfnamefont {A.~M.}\ \bibnamefont {Zheltikov}}, \ and\
  \bibinfo {author} {\bibfnamefont {A.}~\bibnamefont
  {Baltu\ifmmode~\check{s}\else \v{s}\fi{}ka}},\ }\href {\doibase
  10.1103/PhysRevLett.106.147401} {\bibfield  {journal} {\bibinfo  {journal}
  {Phys. Rev. Lett.}\ }\textbf {\bibinfo {volume} {106}},\ \bibinfo {pages}
  {147401} (\bibinfo {year} {2011})}\BibitemShut {NoStop}%
\bibitem [{\citenamefont {Ghimire}\ \emph
  {et~al.}(2011{\natexlab{a}})\citenamefont {Ghimire}, \citenamefont
  {DiChiara}, \citenamefont {Sistrunk}, \citenamefont {Agostini}, \citenamefont
  {DiMauro},\ and\ \citenamefont {Reis}}]{Ghimire_2011_NP_7_138}%
  \BibitemOpen
  \bibfield  {author} {\bibinfo {author} {\bibfnamefont {S.}~\bibnamefont
  {Ghimire}}, \bibinfo {author} {\bibfnamefont {A.~D.}\ \bibnamefont
  {DiChiara}}, \bibinfo {author} {\bibfnamefont {E.}~\bibnamefont {Sistrunk}},
  \bibinfo {author} {\bibfnamefont {P.}~\bibnamefont {Agostini}}, \bibinfo
  {author} {\bibfnamefont {L.~F.}\ \bibnamefont {DiMauro}}, \ and\ \bibinfo
  {author} {\bibfnamefont {D.~A.}\ \bibnamefont {Reis}},\ }\href {\doibase
  10.1038/NPHYS1847} {\bibfield  {journal} {\bibinfo  {journal} {Nature
  Physics}\ }\textbf {\bibinfo {volume} {7}},\ \bibinfo {pages} {138} (\bibinfo
  {year} {2011}{\natexlab{a}})}\BibitemShut {NoStop}%
\bibitem [{\citenamefont {Ghimire}\ \emph
  {et~al.}(2011{\natexlab{b}})\citenamefont {Ghimire}, \citenamefont
  {DiChiara}, \citenamefont {Sistrunk}, \citenamefont {Szafruga}, \citenamefont
  {Agostini}, \citenamefont {DiMauro},\ and\ \citenamefont
  {Reis}}]{Ghimire_2011_PRL_107_167407}%
  \BibitemOpen
  \bibfield  {author} {\bibinfo {author} {\bibfnamefont {S.}~\bibnamefont
  {Ghimire}}, \bibinfo {author} {\bibfnamefont {A.~D.}\ \bibnamefont
  {DiChiara}}, \bibinfo {author} {\bibfnamefont {E.}~\bibnamefont {Sistrunk}},
  \bibinfo {author} {\bibfnamefont {U.~B.}\ \bibnamefont {Szafruga}}, \bibinfo
  {author} {\bibfnamefont {P.}~\bibnamefont {Agostini}}, \bibinfo {author}
  {\bibfnamefont {L.~F.}\ \bibnamefont {DiMauro}}, \ and\ \bibinfo {author}
  {\bibfnamefont {D.~A.}\ \bibnamefont {Reis}},\ }\href {\doibase
  10.1103/PhysRevLett.107.167407} {\bibfield  {journal} {\bibinfo  {journal}
  {Phys. Rev. Lett.}\ }\textbf {\bibinfo {volume} {107}},\ \bibinfo {pages}
  {167407} (\bibinfo {year} {2011}{\natexlab{b}})}\BibitemShut {NoStop}%
\bibitem [{\citenamefont {Schiffrin}\ \emph {et~al.}(2013)\citenamefont
  {Schiffrin}, \citenamefont {Paasch-Colberg}, \citenamefont {Karpowicz},
  \citenamefont {Apalkov}, \citenamefont {Gerster}, \citenamefont
  {M\"uhlbrandt}, \citenamefont {Korbman}, \citenamefont {Reichert},
  \citenamefont {Schultze}, \citenamefont {Holzner}, \citenamefont {Barth},
  \citenamefont {Kienberger}, \citenamefont {Ernstorfer}, \citenamefont
  {Yakovlev}, \citenamefont {Stockman},\ and\ \citenamefont
  {Krausz}}]{Schiffrin_2013_Nature_493_70}%
  \BibitemOpen
  \bibfield  {author} {\bibinfo {author} {\bibfnamefont {A.}~\bibnamefont
  {Schiffrin}}, \bibinfo {author} {\bibfnamefont {T.}~\bibnamefont
  {Paasch-Colberg}}, \bibinfo {author} {\bibfnamefont {N.}~\bibnamefont
  {Karpowicz}}, \bibinfo {author} {\bibfnamefont {V.}~\bibnamefont {Apalkov}},
  \bibinfo {author} {\bibfnamefont {D.}~\bibnamefont {Gerster}}, \bibinfo
  {author} {\bibfnamefont {S.}~\bibnamefont {M\"uhlbrandt}}, \bibinfo {author}
  {\bibfnamefont {M.}~\bibnamefont {Korbman}}, \bibinfo {author} {\bibfnamefont
  {J.}~\bibnamefont {Reichert}}, \bibinfo {author} {\bibfnamefont
  {M.}~\bibnamefont {Schultze}}, \bibinfo {author} {\bibfnamefont
  {S.}~\bibnamefont {Holzner}}, \bibinfo {author} {\bibfnamefont {J.~V.}\
  \bibnamefont {Barth}}, \bibinfo {author} {\bibfnamefont {R.}~\bibnamefont
  {Kienberger}}, \bibinfo {author} {\bibfnamefont {R.}~\bibnamefont
  {Ernstorfer}}, \bibinfo {author} {\bibfnamefont {V.~S.}\ \bibnamefont
  {Yakovlev}}, \bibinfo {author} {\bibfnamefont {M.~I.}\ \bibnamefont
  {Stockman}}, \ and\ \bibinfo {author} {\bibfnamefont {F.}~\bibnamefont
  {Krausz}},\ }\href@noop {} {\bibfield  {journal} {\bibinfo  {journal}
  {Nature}\ }\textbf {\bibinfo {volume} {493}},\ \bibinfo {pages} {70}
  (\bibinfo {year} {2013})}\BibitemShut {NoStop}%
\bibitem [{\citenamefont {Kurizki}\ \emph {et~al.}(1989)\citenamefont
  {Kurizki}, \citenamefont {Shapiro},\ and\ \citenamefont
  {Brumer}}]{Kurizki_1989_PRB_39_3435}%
  \BibitemOpen
  \bibfield  {author} {\bibinfo {author} {\bibfnamefont {G.}~\bibnamefont
  {Kurizki}}, \bibinfo {author} {\bibfnamefont {M.}~\bibnamefont {Shapiro}}, \
  and\ \bibinfo {author} {\bibfnamefont {P.}~\bibnamefont {Brumer}},\ }\href
  {\doibase 10.1103/PhysRevB.39.3435} {\bibfield  {journal} {\bibinfo
  {journal} {Phys. Rev. B}\ }\textbf {\bibinfo {volume} {39}},\ \bibinfo
  {pages} {3435} (\bibinfo {year} {1989})}\BibitemShut {NoStop}%
\bibitem [{\citenamefont {Atanasov}\ \emph {et~al.}(1996)\citenamefont
  {Atanasov}, \citenamefont {Hach\'e}, \citenamefont {Hughes}, \citenamefont
  {van Driel},\ and\ \citenamefont {Sipe}}]{Atanasov_1996_PRL_76_1703}%
  \BibitemOpen
  \bibfield  {author} {\bibinfo {author} {\bibfnamefont {R.}~\bibnamefont
  {Atanasov}}, \bibinfo {author} {\bibfnamefont {A.}~\bibnamefont {Hach\'e}},
  \bibinfo {author} {\bibfnamefont {J.~L.~P.}\ \bibnamefont {Hughes}}, \bibinfo
  {author} {\bibfnamefont {H.~M.}\ \bibnamefont {van Driel}}, \ and\ \bibinfo
  {author} {\bibfnamefont {J.~E.}\ \bibnamefont {Sipe}},\ }\href {\doibase
  10.1103/PhysRevLett.76.1703} {\bibfield  {journal} {\bibinfo  {journal}
  {Phys. Rev. Lett.}\ }\textbf {\bibinfo {volume} {76}},\ \bibinfo {pages}
  {1703} (\bibinfo {year} {1996})}\BibitemShut {NoStop}%
\bibitem [{\citenamefont {Hach\'e}\ \emph {et~al.}(1997)\citenamefont
  {Hach\'e}, \citenamefont {Kostoulas}, \citenamefont {Atanasov}, \citenamefont
  {Hughes}, \citenamefont {Sipe},\ and\ \citenamefont {van
  Driel}}]{Hache_1997_PRL_78_306}%
  \BibitemOpen
  \bibfield  {author} {\bibinfo {author} {\bibfnamefont {A.}~\bibnamefont
  {Hach\'e}}, \bibinfo {author} {\bibfnamefont {Y.}~\bibnamefont {Kostoulas}},
  \bibinfo {author} {\bibfnamefont {R.}~\bibnamefont {Atanasov}}, \bibinfo
  {author} {\bibfnamefont {J.~L.~P.}\ \bibnamefont {Hughes}}, \bibinfo {author}
  {\bibfnamefont {J.~E.}\ \bibnamefont {Sipe}}, \ and\ \bibinfo {author}
  {\bibfnamefont {H.~M.}\ \bibnamefont {van Driel}},\ }\href {\doibase
  10.1103/PhysRevLett.78.306} {\bibfield  {journal} {\bibinfo  {journal} {Phys.
  Rev. Lett.}\ }\textbf {\bibinfo {volume} {78}},\ \bibinfo {pages} {306}
  (\bibinfo {year} {1997})}\BibitemShut {NoStop}%
\bibitem [{\citenamefont {Rioux}\ and\ \citenamefont
  {Sipe}(2012)}]{Rioux_2012_PE_45_1}%
  \BibitemOpen
  \bibfield  {author} {\bibinfo {author} {\bibfnamefont {J.}~\bibnamefont
  {Rioux}}\ and\ \bibinfo {author} {\bibfnamefont {J.}~\bibnamefont {Sipe}},\
  }\href {\doibase 10.1016/j.physe.2012.07.004} {\bibfield  {journal} {\bibinfo
   {journal} {Physica E}\ }\textbf {\bibinfo {volume} {45}},\ \bibinfo {pages}
  {1} (\bibinfo {year} {2012})}\BibitemShut {NoStop}%
\bibitem [{\citenamefont {Nguyen-Dang}\ \emph {et~al.}(2005)\citenamefont
  {Nguyen-Dang}, \citenamefont {Lefebvre}, \citenamefont {Abou-Rachid},\ and\
  \citenamefont {Atabek}}]{Nguyen-Dang_2005_PRA_71_023403}%
  \BibitemOpen
  \bibfield  {author} {\bibinfo {author} {\bibfnamefont {T.~T.}\ \bibnamefont
  {Nguyen-Dang}}, \bibinfo {author} {\bibfnamefont {C.}~\bibnamefont
  {Lefebvre}}, \bibinfo {author} {\bibfnamefont {H.}~\bibnamefont
  {Abou-Rachid}}, \ and\ \bibinfo {author} {\bibfnamefont {O.}~\bibnamefont
  {Atabek}},\ }\href {\doibase 10.1103/PhysRevA.71.023403} {\bibfield
  {journal} {\bibinfo  {journal} {Phys. Rev. A}\ }\textbf {\bibinfo {volume}
  {71}},\ \bibinfo {pages} {023403} (\bibinfo {year} {2005})}\BibitemShut
  {NoStop}%
\bibitem [{\citenamefont {Nakajima}\ and\ \citenamefont
  {Watanabe}(2006)}]{Nakajima_2006_PRL_96_213001}%
  \BibitemOpen
  \bibfield  {author} {\bibinfo {author} {\bibfnamefont {T.}~\bibnamefont
  {Nakajima}}\ and\ \bibinfo {author} {\bibfnamefont {S.}~\bibnamefont
  {Watanabe}},\ }\href {\doibase 10.1103/PhysRevLett.96.213001} {\bibfield
  {journal} {\bibinfo  {journal} {Phys. Rev. Lett.}\ }\textbf {\bibinfo
  {volume} {96}},\ \bibinfo {pages} {213001} (\bibinfo {year}
  {2006})}\BibitemShut {NoStop}%
\bibitem [{\citenamefont {Roudnev}\ and\ \citenamefont
  {Esry}(2007)}]{Roudnev_2007_PRL_99_220406}%
  \BibitemOpen
  \bibfield  {author} {\bibinfo {author} {\bibfnamefont {V.}~\bibnamefont
  {Roudnev}}\ and\ \bibinfo {author} {\bibfnamefont {B.~D.}\ \bibnamefont
  {Esry}},\ }\href {\doibase 10.1103/PhysRevLett.99.220406} {\bibfield
  {journal} {\bibinfo  {journal} {Phys. Rev. Lett.}\ }\textbf {\bibinfo
  {volume} {99}},\ \bibinfo {pages} {220406} (\bibinfo {year}
  {2007})}\BibitemShut {NoStop}%
\bibitem [{\citenamefont {Haug}\ and\ \citenamefont {Koch}(2004)}]{Haug_2004}%
  \BibitemOpen
  \bibfield  {author} {\bibinfo {author} {\bibfnamefont {H.}~\bibnamefont
  {Haug}}\ and\ \bibinfo {author} {\bibfnamefont {S.~W.}\ \bibnamefont
  {Koch}},\ }\href@noop {} {\emph {\bibinfo {title} {Quantum theory of the
  optical and electronic properties of semiconductors}}}\ (\bibinfo
  {publisher} {World Scientific},\ \bibinfo {year} {2004})\BibitemShut
  {NoStop}%
\bibitem [{\citenamefont {Rossi}\ and\ \citenamefont
  {Kuhn}(2002)}]{Rossi_2002_RMP_74_895}%
  \BibitemOpen
  \bibfield  {author} {\bibinfo {author} {\bibfnamefont {F.}~\bibnamefont
  {Rossi}}\ and\ \bibinfo {author} {\bibfnamefont {T.}~\bibnamefont {Kuhn}},\
  }\href {\doibase 10.1103/RevModPhys.74.895} {\bibfield  {journal} {\bibinfo
  {journal} {Rev. Mod. Phys.}\ }\textbf {\bibinfo {volume} {74}},\ \bibinfo
  {pages} {895} (\bibinfo {year} {2002})}\BibitemShut {NoStop}%
\bibitem [{\citenamefont {Bachau}\ \emph {et~al.}(2006)\citenamefont {Bachau},
  \citenamefont {Belsky}, \citenamefont {Martin}, \citenamefont {Vasil'ev},\
  and\ \citenamefont {Yatsenko}}]{Bachau_2006_PRB_74_235215}%
  \BibitemOpen
  \bibfield  {author} {\bibinfo {author} {\bibfnamefont {H.}~\bibnamefont
  {Bachau}}, \bibinfo {author} {\bibfnamefont {A.~N.}\ \bibnamefont {Belsky}},
  \bibinfo {author} {\bibfnamefont {P.}~\bibnamefont {Martin}}, \bibinfo
  {author} {\bibfnamefont {A.~N.}\ \bibnamefont {Vasil'ev}}, \ and\ \bibinfo
  {author} {\bibfnamefont {B.~N.}\ \bibnamefont {Yatsenko}},\ }\href {\doibase
  10.1103/PhysRevB.74.235215} {\bibfield  {journal} {\bibinfo  {journal} {Phys.
  Rev. B}\ }\textbf {\bibinfo {volume} {74}},\ \bibinfo {pages} {235215}
  (\bibinfo {year} {2006})}\BibitemShut {NoStop}%
\bibitem [{\citenamefont {Korbman}\ \emph {et~al.}(2013)\citenamefont
  {Korbman}, \citenamefont {Kruchinin},\ and\ \citenamefont
  {Yakovlev}}]{Korbman_2013_NJP_15_013006}%
  \BibitemOpen
  \bibfield  {author} {\bibinfo {author} {\bibfnamefont {M.}~\bibnamefont
  {Korbman}}, \bibinfo {author} {\bibfnamefont {S.~{\relax Yu.}.}\ \bibnamefont
  {Kruchinin}}, \ and\ \bibinfo {author} {\bibfnamefont {V.~S.}\ \bibnamefont
  {Yakovlev}},\ }\href@noop {} {\bibfield  {journal} {\bibinfo  {journal} {New
  J. Phys.}\ }\textbf {\bibinfo {volume} {15}},\ \bibinfo {pages} {013006}
  (\bibinfo {year} {2013})}\BibitemShut {NoStop}%
\bibitem [{\citenamefont {Axt}\ and\ \citenamefont
  {Kuhn}(2004)}]{Axt_2004_RPP_67_433}%
  \BibitemOpen
  \bibfield  {author} {\bibinfo {author} {\bibfnamefont {V.~M.}\ \bibnamefont
  {Axt}}\ and\ \bibinfo {author} {\bibfnamefont {T.}~\bibnamefont {Kuhn}},\
  }\href@noop {} {\bibfield  {journal} {\bibinfo  {journal} {Rep. Prog. Phys.}\
  }\textbf {\bibinfo {volume} {67}},\ \bibinfo {pages} {433} (\bibinfo {year}
  {2004})}\BibitemShut {NoStop}%
\bibitem [{\citenamefont {Datta}(1995)}]{Datta_1995}%
  \BibitemOpen
  \bibfield  {author} {\bibinfo {author} {\bibfnamefont {S.}~\bibnamefont
  {Datta}},\ }\href@noop {} {\emph {\bibinfo {title} {Electronic Transport in
  Mesoscopic Systems}}}\ (\bibinfo  {publisher} {Cambridge University Press},\
  \bibinfo {year} {1995})\BibitemShut {NoStop}%
\bibitem [{\citenamefont {Jacobini}(2010)}]{Jacobini_2010}%
  \BibitemOpen
  \bibfield  {author} {\bibinfo {author} {\bibfnamefont {C.}~\bibnamefont
  {Jacobini}},\ }\href@noop {} {\emph {\bibinfo {title} {Theory of Electronic
  Transport in Semiconductors: A pathway from Elementary Physics to
  Nonequilibrium Green Functions}}}\ (\bibinfo  {publisher} {Springer},\
  \bibinfo {year} {2010})\BibitemShut {NoStop}%
\bibitem [{\citenamefont {Bransden}\ and\ \citenamefont
  {Joachain}(2000)}]{Bransden_2000}%
  \BibitemOpen
  \bibfield  {author} {\bibinfo {author} {\bibfnamefont {B.~H.}\ \bibnamefont
  {Bransden}}\ and\ \bibinfo {author} {\bibfnamefont {C.~J.}\ \bibnamefont
  {Joachain}},\ }\href@noop {} {\emph {\bibinfo {title} {Quantum Mechanics}}}\
  (\bibinfo  {publisher} {Pearson Education Limited},\ \bibinfo {year}
  {2000})\BibitemShut {NoStop}%
\bibitem [{\citenamefont {Blount}(1962)}]{Blount_1962}%
  \BibitemOpen
  \bibfield  {author} {\bibinfo {author} {\bibfnamefont {E.~I.}\ \bibnamefont
  {Blount}}\ }(\bibinfo  {publisher} {Academic Press},\ \bibinfo {year}
  {1962})\ pp.\ \bibinfo {pages} {305--373}\BibitemShut {NoStop}%
\bibitem [{\citenamefont {Aversa}\ and\ \citenamefont
  {Sipe}(1995)}]{Aversa_1995_PRB_52_14636}%
  \BibitemOpen
  \bibfield  {author} {\bibinfo {author} {\bibfnamefont {C.}~\bibnamefont
  {Aversa}}\ and\ \bibinfo {author} {\bibfnamefont {J.~E.}\ \bibnamefont
  {Sipe}},\ }\href {\doibase 10.1103/PhysRevB.52.14636} {\bibfield  {journal}
  {\bibinfo  {journal} {Phys. Rev. B}\ }\textbf {\bibinfo {volume} {52}},\
  \bibinfo {pages} {14636} (\bibinfo {year} {1995})}\BibitemShut {NoStop}%
\bibitem [{\citenamefont {Virk}\ and\ \citenamefont
  {Sipe}(2007)}]{Virk_2007_PRB_76_035213}%
  \BibitemOpen
  \bibfield  {author} {\bibinfo {author} {\bibfnamefont {K.~S.}\ \bibnamefont
  {Virk}}\ and\ \bibinfo {author} {\bibfnamefont {J.~E.}\ \bibnamefont
  {Sipe}},\ }\href {\doibase 10.1103/PhysRevB.76.035213} {\bibfield  {journal}
  {\bibinfo  {journal} {Phys. Rev. B}\ }\textbf {\bibinfo {volume} {76}},\
  \bibinfo {pages} {035213} (\bibinfo {year} {2007})}\BibitemShut {NoStop}%
\bibitem [{\citenamefont {Otobe}\ \emph {et~al.}(2008)\citenamefont {Otobe},
  \citenamefont {Yamagiwa}, \citenamefont {Iwata}, \citenamefont {Yabana},
  \citenamefont {Nakatsukasa},\ and\ \citenamefont
  {Bertsch}}]{Otobe_2008_PRB_77_165104}%
  \BibitemOpen
  \bibfield  {author} {\bibinfo {author} {\bibfnamefont {T.}~\bibnamefont
  {Otobe}}, \bibinfo {author} {\bibfnamefont {M.}~\bibnamefont {Yamagiwa}},
  \bibinfo {author} {\bibfnamefont {J.-I.}\ \bibnamefont {Iwata}}, \bibinfo
  {author} {\bibfnamefont {K.}~\bibnamefont {Yabana}}, \bibinfo {author}
  {\bibfnamefont {T.}~\bibnamefont {Nakatsukasa}}, \ and\ \bibinfo {author}
  {\bibfnamefont {G.~F.}\ \bibnamefont {Bertsch}},\ }\href {\doibase
  10.1103/PhysRevB.77.165104} {\bibfield  {journal} {\bibinfo  {journal} {Phys.
  Rev. B}\ }\textbf {\bibinfo {volume} {77}},\ \bibinfo {pages} {165104}
  (\bibinfo {year} {2008})}\BibitemShut {NoStop}%
\bibitem [{\citenamefont {Yabana}\ \emph {et~al.}(2012)\citenamefont {Yabana},
  \citenamefont {Sugiyama}, \citenamefont {Shinohara}, \citenamefont {Otobe},\
  and\ \citenamefont {Bertsch}}]{Yabana_2012_PRB_85_045134}%
  \BibitemOpen
  \bibfield  {author} {\bibinfo {author} {\bibfnamefont {K.}~\bibnamefont
  {Yabana}}, \bibinfo {author} {\bibfnamefont {T.}~\bibnamefont {Sugiyama}},
  \bibinfo {author} {\bibfnamefont {Y.}~\bibnamefont {Shinohara}}, \bibinfo
  {author} {\bibfnamefont {T.}~\bibnamefont {Otobe}}, \ and\ \bibinfo {author}
  {\bibfnamefont {G.~F.}\ \bibnamefont {Bertsch}},\ }\href {\doibase
  10.1103/PhysRevB.85.045134} {\bibfield  {journal} {\bibinfo  {journal} {Phys.
  Rev. B}\ }\textbf {\bibinfo {volume} {85}},\ \bibinfo {pages} {045134}
  (\bibinfo {year} {2012})}\BibitemShut {NoStop}%
\bibitem [{\citenamefont {Hess}\ and\ \citenamefont
  {Kuhn}(1996)}]{Hess_1996_PRA_54_3347}%
  \BibitemOpen
  \bibfield  {author} {\bibinfo {author} {\bibfnamefont {O.}~\bibnamefont
  {Hess}}\ and\ \bibinfo {author} {\bibfnamefont {T.}~\bibnamefont {Kuhn}},\
  }\href {\doibase 10.1103/PhysRevA.54.3347} {\bibfield  {journal} {\bibinfo
  {journal} {Phys. Rev. A}\ }\textbf {\bibinfo {volume} {54}},\ \bibinfo
  {pages} {3347} (\bibinfo {year} {1996})}\BibitemShut {NoStop}%
\bibitem [{\citenamefont {Giessen}\ \emph {et~al.}(1998)\citenamefont
  {Giessen}, \citenamefont {Knorr}, \citenamefont {Haas}, \citenamefont {Koch},
  \citenamefont {Linden}, \citenamefont {Kuhl}, \citenamefont {Hetterich},
  \citenamefont {Gr\"un},\ and\ \citenamefont
  {Klingshirn}}]{Giessen_1998_PRL_81_4260}%
  \BibitemOpen
  \bibfield  {author} {\bibinfo {author} {\bibfnamefont {H.}~\bibnamefont
  {Giessen}}, \bibinfo {author} {\bibfnamefont {A.}~\bibnamefont {Knorr}},
  \bibinfo {author} {\bibfnamefont {S.}~\bibnamefont {Haas}}, \bibinfo {author}
  {\bibfnamefont {S.~W.}\ \bibnamefont {Koch}}, \bibinfo {author}
  {\bibfnamefont {S.}~\bibnamefont {Linden}}, \bibinfo {author} {\bibfnamefont
  {J.}~\bibnamefont {Kuhl}}, \bibinfo {author} {\bibfnamefont {M.}~\bibnamefont
  {Hetterich}}, \bibinfo {author} {\bibfnamefont {M.}~\bibnamefont {Gr\"un}}, \
  and\ \bibinfo {author} {\bibfnamefont {C.}~\bibnamefont {Klingshirn}},\
  }\href {\doibase 10.1103/PhysRevLett.81.4260} {\bibfield  {journal} {\bibinfo
   {journal} {Phys. Rev. Lett.}\ }\textbf {\bibinfo {volume} {81}},\ \bibinfo
  {pages} {4260} (\bibinfo {year} {1998})}\BibitemShut {NoStop}%
\bibitem [{\citenamefont {Rammer}(1998)}]{Rammer_1998}%
  \BibitemOpen
  \bibfield  {author} {\bibinfo {author} {\bibfnamefont {J.}~\bibnamefont
  {Rammer}},\ }\href@noop {} {\emph {\bibinfo {title} {Quantum Transport
  Theory}}}\ (\bibinfo  {publisher} {Perseus Books},\ \bibinfo {year}
  {1998})\BibitemShut {NoStop}%
\bibitem [{\citenamefont {Cohen}\ and\ \citenamefont
  {Chelikowsky}(1988)}]{Cohen_1988}%
  \BibitemOpen
  \bibfield  {author} {\bibinfo {author} {\bibfnamefont {M.~L.}\ \bibnamefont
  {Cohen}}\ and\ \bibinfo {author} {\bibfnamefont {J.~R.}\ \bibnamefont
  {Chelikowsky}},\ }\href@noop {} {\emph {\bibinfo {title} {Electronic
  Structure and Optical Properties of Semiconductors}}}\ (\bibinfo  {publisher}
  {Springer},\ \bibinfo {year} {1988})\BibitemShut {NoStop}%
\bibitem [{\citenamefont {Schneider}\ and\ \citenamefont
  {Fowler}(1976)}]{Schneider_1976_PRL_36_425}%
  \BibitemOpen
  \bibfield  {author} {\bibinfo {author} {\bibfnamefont {P.~M.}\ \bibnamefont
  {Schneider}}\ and\ \bibinfo {author} {\bibfnamefont {W.~B.}\ \bibnamefont
  {Fowler}},\ }\href {\doibase 10.1103/PhysRevLett.36.425} {\bibfield
  {journal} {\bibinfo  {journal} {Phys. Rev. Lett.}\ }\textbf {\bibinfo
  {volume} {36}},\ \bibinfo {pages} {425} (\bibinfo {year} {1976})}\BibitemShut
  {NoStop}%
\bibitem [{\citenamefont {Chelikowsky}\ and\ \citenamefont
  {Schl\"uter}(1977)}]{Chelikowsky_1977_PRB_15_4020}%
  \BibitemOpen
  \bibfield  {author} {\bibinfo {author} {\bibfnamefont {J.~R.}\ \bibnamefont
  {Chelikowsky}}\ and\ \bibinfo {author} {\bibfnamefont {M.}~\bibnamefont
  {Schl\"uter}},\ }\href {\doibase 10.1103/PhysRevB.15.4020} {\bibfield
  {journal} {\bibinfo  {journal} {Phys. Rev. B}\ }\textbf {\bibinfo {volume}
  {15}},\ \bibinfo {pages} {4020} (\bibinfo {year} {1977})}\BibitemShut
  {NoStop}%
\bibitem [{\citenamefont {Gnani}\ \emph {et~al.}(2000)\citenamefont {Gnani},
  \citenamefont {Reggiani}, \citenamefont {Colle},\ and\ \citenamefont
  {Rudan}}]{Gnani_2000_ITED_47_1795}%
  \BibitemOpen
  \bibfield  {author} {\bibinfo {author} {\bibfnamefont {E.}~\bibnamefont
  {Gnani}}, \bibinfo {author} {\bibfnamefont {S.}~\bibnamefont {Reggiani}},
  \bibinfo {author} {\bibfnamefont {R.}~\bibnamefont {Colle}}, \ and\ \bibinfo
  {author} {\bibfnamefont {M.}~\bibnamefont {Rudan}},\ }\href {\doibase
  10.1109/16.870550} {\bibfield  {journal} {\bibinfo  {journal} {IEEE
  Transactions on Electron Devices}\ }\textbf {\bibinfo {volume} {47}},\
  \bibinfo {pages} {1795} (\bibinfo {year} {2000})}\BibitemShut {NoStop}%
\bibitem [{\citenamefont {Fortier}\ \emph {et~al.}(2004)\citenamefont
  {Fortier}, \citenamefont {Roos}, \citenamefont {Jones}, \citenamefont
  {Cundiff}, \citenamefont {Bhat},\ and\ \citenamefont
  {Sipe}}]{Fortier_2004_PRL_92_147403}%
  \BibitemOpen
  \bibfield  {author} {\bibinfo {author} {\bibfnamefont {T.~M.}\ \bibnamefont
  {Fortier}}, \bibinfo {author} {\bibfnamefont {P.~A.}\ \bibnamefont {Roos}},
  \bibinfo {author} {\bibfnamefont {D.~J.}\ \bibnamefont {Jones}}, \bibinfo
  {author} {\bibfnamefont {S.~T.}\ \bibnamefont {Cundiff}}, \bibinfo {author}
  {\bibfnamefont {R.~D.~R.}\ \bibnamefont {Bhat}}, \ and\ \bibinfo {author}
  {\bibfnamefont {J.~E.}\ \bibnamefont {Sipe}},\ }\href {\doibase
  10.1103/PhysRevLett.92.147403} {\bibfield  {journal} {\bibinfo  {journal}
  {Phys. Rev. Lett.}\ }\textbf {\bibinfo {volume} {92}},\ \bibinfo {pages}
  {147403} (\bibinfo {year} {2004})}\BibitemShut {NoStop}%
\bibitem [{\citenamefont {Kersting}\ \emph {et~al.}(1997)\citenamefont
  {Kersting}, \citenamefont {Unterrainer}, \citenamefont {Strasser},
  \citenamefont {Kauffmann},\ and\ \citenamefont
  {Gornik}}]{Kersting_1997_PRL_79_3038}%
  \BibitemOpen
  \bibfield  {author} {\bibinfo {author} {\bibfnamefont {R.}~\bibnamefont
  {Kersting}}, \bibinfo {author} {\bibfnamefont {K.}~\bibnamefont
  {Unterrainer}}, \bibinfo {author} {\bibfnamefont {G.}~\bibnamefont
  {Strasser}}, \bibinfo {author} {\bibfnamefont {H.~F.}\ \bibnamefont
  {Kauffmann}}, \ and\ \bibinfo {author} {\bibfnamefont {E.}~\bibnamefont
  {Gornik}},\ }\href {\doibase 10.1103/PhysRevLett.79.3038} {\bibfield
  {journal} {\bibinfo  {journal} {Phys. Rev. Lett.}\ }\textbf {\bibinfo
  {volume} {79}},\ \bibinfo {pages} {3038} (\bibinfo {year}
  {1997})}\BibitemShut {NoStop}%
\bibitem [{\citenamefont {Bonitz}\ \emph {et~al.}(2000)\citenamefont {Bonitz},
  \citenamefont {Lampin}, \citenamefont {Camescasse},\ and\ \citenamefont
  {Alexandrou}}]{Bonitz_2000_PRB_62_15724}%
  \BibitemOpen
  \bibfield  {author} {\bibinfo {author} {\bibfnamefont {M.}~\bibnamefont
  {Bonitz}}, \bibinfo {author} {\bibfnamefont {J.~F.}\ \bibnamefont {Lampin}},
  \bibinfo {author} {\bibfnamefont {F.~X.}\ \bibnamefont {Camescasse}}, \ and\
  \bibinfo {author} {\bibfnamefont {A.}~\bibnamefont {Alexandrou}},\ }\href
  {\doibase 10.1103/PhysRevB.62.15724} {\bibfield  {journal} {\bibinfo
  {journal} {Phys. Rev. B}\ }\textbf {\bibinfo {volume} {62}},\ \bibinfo
  {pages} {15724} (\bibinfo {year} {2000})}\BibitemShut {NoStop}%
\bibitem [{\citenamefont {Meinert}\ \emph {et~al.}(2000)\citenamefont
  {Meinert}, \citenamefont {B\'anyai}, \citenamefont {Gartner},\ and\
  \citenamefont {Haug}}]{Meinert_2000_PRB_62_5003}%
  \BibitemOpen
  \bibfield  {author} {\bibinfo {author} {\bibfnamefont {G.}~\bibnamefont
  {Meinert}}, \bibinfo {author} {\bibfnamefont {L.}~\bibnamefont {B\'anyai}},
  \bibinfo {author} {\bibfnamefont {P.}~\bibnamefont {Gartner}}, \ and\
  \bibinfo {author} {\bibfnamefont {H.}~\bibnamefont {Haug}},\ }\href {\doibase
  10.1103/PhysRevB.62.5003} {\bibfield  {journal} {\bibinfo  {journal} {Phys.
  Rev. B}\ }\textbf {\bibinfo {volume} {62}},\ \bibinfo {pages} {5003}
  (\bibinfo {year} {2000})}\BibitemShut {NoStop}%
\bibitem [{\citenamefont {Bass}\ \emph {et~al.}(1962)\citenamefont {Bass},
  \citenamefont {Franken}, \citenamefont {Ward},\ and\ \citenamefont
  {Weinreich}}]{Bass_1962_PRL_9_446}%
  \BibitemOpen
  \bibfield  {author} {\bibinfo {author} {\bibfnamefont {M.}~\bibnamefont
  {Bass}}, \bibinfo {author} {\bibfnamefont {P.~A.}\ \bibnamefont {Franken}},
  \bibinfo {author} {\bibfnamefont {J.~F.}\ \bibnamefont {Ward}}, \ and\
  \bibinfo {author} {\bibfnamefont {G.}~\bibnamefont {Weinreich}},\ }\href
  {\doibase 10.1103/PhysRevLett.9.446} {\bibfield  {journal} {\bibinfo
  {journal} {Phys. Rev. Lett.}\ }\textbf {\bibinfo {volume} {9}},\ \bibinfo
  {pages} {446} (\bibinfo {year} {1962})}\BibitemShut {NoStop}%
\bibitem [{\citenamefont {Rice}\ \emph {et~al.}(1994)\citenamefont {Rice},
  \citenamefont {Jin}, \citenamefont {Ma}, \citenamefont {Zhang}, \citenamefont
  {Bliss}, \citenamefont {Larkin},\ and\ \citenamefont
  {Alexander}}]{Rice_1994_APL_64_1324}%
  \BibitemOpen
  \bibfield  {author} {\bibinfo {author} {\bibfnamefont {A.}~\bibnamefont
  {Rice}}, \bibinfo {author} {\bibfnamefont {Y.}~\bibnamefont {Jin}}, \bibinfo
  {author} {\bibfnamefont {X.~F.}\ \bibnamefont {Ma}}, \bibinfo {author}
  {\bibfnamefont {X.-C.}\ \bibnamefont {Zhang}}, \bibinfo {author}
  {\bibfnamefont {D.}~\bibnamefont {Bliss}}, \bibinfo {author} {\bibfnamefont
  {J.}~\bibnamefont {Larkin}}, \ and\ \bibinfo {author} {\bibfnamefont
  {M.}~\bibnamefont {Alexander}},\ }\href {\doibase 10.1063/1.111922}
  {\bibfield  {journal} {\bibinfo  {journal} {Applied Physics Letters}\
  }\textbf {\bibinfo {volume} {64}},\ \bibinfo {pages} {1324} (\bibinfo {year}
  {1994})}\BibitemShut {NoStop}%
\end{thebibliography}%
\end{document}